\documentclass[fleqn,usenatbib]{mnras}

\usepackage{newtxtext,newtxmath}

\usepackage[T1]{fontenc}

\DeclareRobustCommand{\VAN}[3]{#2}
\let\VANthebibliography\thebibliography
\def\thebibliography{\DeclareRobustCommand{\VAN}[3]{##3}\VANthebibliography}

\usepackage{graphicx}	
\usepackage{caption}            %
\usepackage{subcaption} 
\usepackage{cite}
\usepackage{amsmath}	
\usepackage{booktabs}
\usepackage{siunitx}
\AtBeginDocument{\RenewCommandCopy\qty\SI}
\RenewCommandCopy\qty\SI
\ExplSyntaxOn
\msg_redirect_name:nnn { siunitx } { physics-pkg } { none }
\ExplSyntaxOff
\usepackage{multirow}   
\usepackage{hyperref}   
\usepackage[capitalise]{cleveref}   
\usepackage{soul}

\newcommand\MBH{\hbox{$M_\text{BH}$}} 
\newcommand\Msun{\hbox{ M$_\odot$}} 

\newcommand{\ie}{i.e.\ } 
\DeclareUnicodeCharacter{2609}{\odot}
\DeclareUnicodeCharacter{2605}{\star}

\title[Searching for IMBH in compact stellar systems]{A search for intermediate-mass black holes in compact stellar systems through optical emissions from tidal disruption events}

\author[R. T. Pomeroy \& M. A. Norris]{
Richard T. Pomeroy,$^{1,2}$
Mark A. Norris,$^{1}$\thanks{E-mail: mnorris2@uclan.ac.uk}
\\
$^{1}$Jeremiah Horrocks Institute, University of Central Lancashire, Preston PR1 2HE, UK\\
$^{2}$Department of Physics and Astronomy, The University of Texas Rio Grande Valley, Brownsville, TX 78520, USA\\
}

\date{Accepted: 2024 Apr 4; Revised: 2024 April 3; Received: 2023 November 28}

\pubyear{2024}

\begin{document}
\label{firstpage}
\pagerange{\pageref{firstpage}--\pageref{lastpage}}
\maketitle

\begin{abstract}
Intermediate-mass black holes (IMBH) are expected to exist in globular clusters (GCs) and compact stellar systems (CSS) in general, but none have been conclusively detected. Tidal disruption events (TDEs), where a star is tidally disrupted by the gravitational field of a black hole, have been observed to occur around the supermassive black holes (SMBH) found at the centres of galaxies, and should also arise around IMBHs, especially in the dense stellar cores of CSS's. However, to date none have been observed in such environments.  Using data from the Zwicky Transient Facility (ZTF) we search for TDEs associated with CSS, but none are found. This non-detection allows us to set an upper limit on the TDE rate in CSS of $n_\text{TDE,Total}\lessapprox10^{-7} CSS^{-1}\text{yr}^{-1}$ which is two dex. below the observed TDE rate involving SMBH interacting with 1\Msun\ main sequence stars in the nuclei of massive galaxies. We also consider ultra compact dwarfs (UCDs) formed through a tidal stripping process in the surveyed volume. On the assumption these CSS contain SMBH and TDE rates are comparable to current observed optical rates in galactic nuclei ($\approx\num{3.2E-5}\text{gal}^{-1}\text{yr}^{-1}$), we determine an upper limit for the number of UCDs formed through a tidal stripping process in the surveyed volume to be $N_\text{GC, Strip}<\num{1.4E4}$, which we estimate represents $< 6\%$ of the population of GCs $>10^6\Msun$.

\end{abstract}




\section{Introduction}\label{intro}

Tidal disruption events occur when the gravitational field of a black hole (BH) overcomes the internal gravity of a star which approaches the BH's tidal radius, $R_\text{T}$. Early theoretical studies \citep[e.g.][]{Frank_1976, Evans_1989} suggested \textasciitilde50\% of the stellar material is unbound and ejects into the ISM, but the remaining material settles into an accretion disc to be consumed by the BH. Spectral profile, intensity and duration of emission from the accretion disc vary greatly depending upon the impact parameters of the encounter, the disrupted star structure, mass and BH mass \citep[e.g.][]{Guillochon_2013,Malyali_2019}. While the rate of discovery of TDEs has increased at all wavelengths since the first candidate observation in X-rays \citep{Bade_1996}, UV \citep{Gezari_2006}, $\gamma$-rays \citep{Bloom_2011} and optical \citep{vanVelzen_2011}, optical observations have dominated recent discoveries \citep[e.g.][]{Gezari_2021}. The increase in optical detections in the last decade is due to the introduction of high cadence optical surveys, regularly revisiting wide areas of sky. Optical facilities involved in TDE detection include SDSS \citep{Frieman_2008}, PanSTARRS \citep{Chambers_2016}, PTF/iPTF \citep{Law_2009, Masci_2017}, ASAS-SN \citep{Kochanek_2017}, OGLE \citep{Udalski_2015}, ATLAS \citep{Tonry_2018}, GAIA \citep{GaiaCollaboration_2016} and ZTF \citep{Bellm_2019a}. 
Even so, TDEs remain rare, but the depth and duration of archive data now permits limits to be placed on TDE rates.

Intermediate-mass black holes (IMBHs) are compact objects, with masses between those of stellar mass and super-massive black holes (SMBH), i.e. $\sim100\Msun\!<\!\MBH\!<\!10^6\Msun$. IMBH are theorised to form in a number of ways \citep[see][]{Greene_2020}, but their formation through mergers in dense stellar clusters \citep[e.g.][]{PortegiesZwart_2002, Miller_2002} is key to why compact stellar systems (CSS) have, to date, been focused on as potential hosts. Merging of BH in dense stellar environments \citep[e.g.][]{Rodriguez_2018, Rodriguez_2019, 
DiCarlo_2021, Gonzalez_2021} has gained renewed interest since the first detection of a gravitational wave event leading to a remnant of IMBH mass \citep{Abbott_2020}. 

The authors also note a recent paper by \citet{Gomez_2023}, who carried out a search for white dwarf (WD) based TDEs, suspected to be associated with IMBH and have similar signatures to Type 1a supernova (SN). These authors did not detect any conclusive events in the ZTF data, based on their specific selection criteria. Similarly, \citet{Tang_2024} searched for TDE events in a sample of nearly 4000 GCs in the Next Generation Virgo Cluster Survey (NGVCS), but did not detect any potential TDEs in their sample.

Within this work, we consider two types of CSS exhibiting high stellar density, which are potentially hosts to IMBH or SMBH: globular clusters (GCs) which dominate our sample by number and ultracompact dwarfs (UCDs). The characteristics of UCDs place them between GCs and and compact ellipticals (cEs) in the mass / size parameter space \citep[see e.g.][]{Norris_2014}. 

GCs generally form at high \textit{z}, exist almost exclusively in galaxy halos and have masses $\lesssim\!10^6 \Msun$, with effective radii $R_e\lesssim10$pc. 



UCDs have masses $\sim\!10^6\! -\! 10^8 \Msun$ and size $10\lesssim\! R_e\lesssim\! 100$ pc. Current research \citep[e.g.][]{Norris_2015, Pfeffer_2016} suggests multiple formation paths where UCDs represent a continuation of the high mass end of GCs, or the tidally stripped remnant of a nucleated dwarf galaxy \citep{Bekki_2001}. Most studies suggest a composite UCD population of the two formation methods \citep[e.g.][]{Norris_2011,Norris_2014, Pfeffer_2016}, with both present below a star cluster formation limit of $M_\star\lesssim\!5\times10^7\Msun$ \citep{Norris_2019}. SMBHs have been detected in 5 (putative stripped nucleus type) UCDs \citep{Seth_2014,Ahn_2017, Ahn_2018, Afanasiev_2018} with upper limits in 2 UCDs determined by \citet{Voggel_2018}. More recently, \citet{Pechetti_2022} suggest the presence of an IMBH $\sim\!10^5\Msun$ in a stripped nucleus UCD around M31. Our search is thus well placed to detect both IMBH and SMBH in UCDs formed as stripped remnants.

The observational data for this analysis was obtained from the Zwicky Transient Facility \citep[ZTF,][]{Bellm_2019b} based around the Palomar 48" Schmidt telescope (Samuel Oschin telescope).
The ZTF is ideally suited to the detection of optical transients from unclassified sources. This is due to the cadence of its public survey observations, where the project aims to complete repeat visits of the entire northern hemisphere every three nights. This allows detection of transients on the order of days, while the dual band observations ($g$ \& $r$) allow colour information to be obtained. To the date of this paper, the ZTF had been credited with the initial discovery of more optical TDE signatures than any other single facility, \ie, 46 of 77 transients classified as TDEs in the Transient Name Server (\href{https://www.wis-tns.org}{TNS}). There is also a good synergy of ZTF with the deeper PanSTARRS data, in terms of checking any off-nuclear TDEs observed for the presence of co-located CSS.

This project aims to add to the area of research which explores potential formation paths of both CSS and IMBH, by searching for TDE signatures of IMBH within the CSS outlined. Detection of new candidate IMBH, or provision of supplementary evidence for the presence of existing candidates, in CSS environments will assist in constraining hypotheses not only of CSS formation, but also of BH formation in dense cluster environments, as detailed above.

This document is structured to outline first the sample data used, including both the host galaxy selection and associated cone search definition (\cref{sample}), followed by details of the light curve extraction from the ZTF archives, through the ALeRCE broker, including filtering and categorisation details (\cref{Alert Analysis}). An analysis of the implications of our findings are outlined in the discussion (\cref{discussion}).

\section{Sample} \label{sample}
Our starting sample was drawn from the Hyperleda database \citep{Makarov_2014}. The criteria for sample selection was defined by the need to follow up any off-nuclear TDEs observed using PanSTARRS data, and to ensure completeness of our host galaxy data. 

The upper limit to GCs has been proposed to be around $M_v = -13$ \citep{Norris_2011, Norris_2019}, and at an upper distance limit of 120 Mpc this equates to an apparent magnitude of $m_v=22.4$ mag, giving a good probability of detection in PanSTARRS with a $5\sigma, g$-band limit of 23.3 mag. The original LEDA database (i.e., the precursor of the Hyperleda catalog) was used as the basis for the Third Reference Catalogue of Bright Galaxies \citep{deVaucouleurs_1991} which was defined as complete to 15.5 mag and has subsequently been augmented with additional mission data to provide the homogenised dataset seen in the Hyperleda database. At this completeness limit for the Hyperleda database, with an upper distance limit of 120 Mpc, gives a distance modulus of 35.4, and an upper limit on the absolute B magnitude of the host galaxies of $M_B$ = -20. Therefore, we expect the input galaxy sample to be close to complete to $M_B$ = -20 within the survey volume. This limit therefore rejects the vast majority of galaxies, however such low mass galaxies have very sparsely populated high-mass GC populations \citep[e.g.][]{vandenBergh_2006} and would tend not to produce GCs of masses where large IMBH’s would be expected \citep[e.g.][]{Lutzgendorf_2013}.

The full criteria used for the sample of CSS host galaxies which could be cross-matched with alerts, was as follows:
\begin{itemize}
    \item $objtype=$ `G': ensure object is defined as a galaxy.
    \item $26.5<modbest<35.4$: best estimate of distance modulus. Gives distance $2\leq D\leq120$ Mpc. The lower limit takes the sample just outside of Andromeda (M31) to reduce clutter and filtering of oversize objects. The upper limit for host galaxies was set as discussed above.
    \item $de2000>-30$: limit on declination $\delta$ imposed by the use of the ZTF facility based in a northerly latitude.
    \item $B\text{mag}<-20$: Using the a threshold on absolute $B$mag ensured rejection of smaller galaxies without significant CSS populations.
\end{itemize}

The initial sample returned included 4812 host galaxies ($A_\text{g}<2.1$ mag). However, to avoid regions of high galactic extinction, a limit of $A_\text{g}<0.3$ mag was imposed, after which 3714 galaxies (\textasciitilde77\%) remained.

\begin{figure}
    \centering
    \includegraphics[width=0.95\columnwidth]{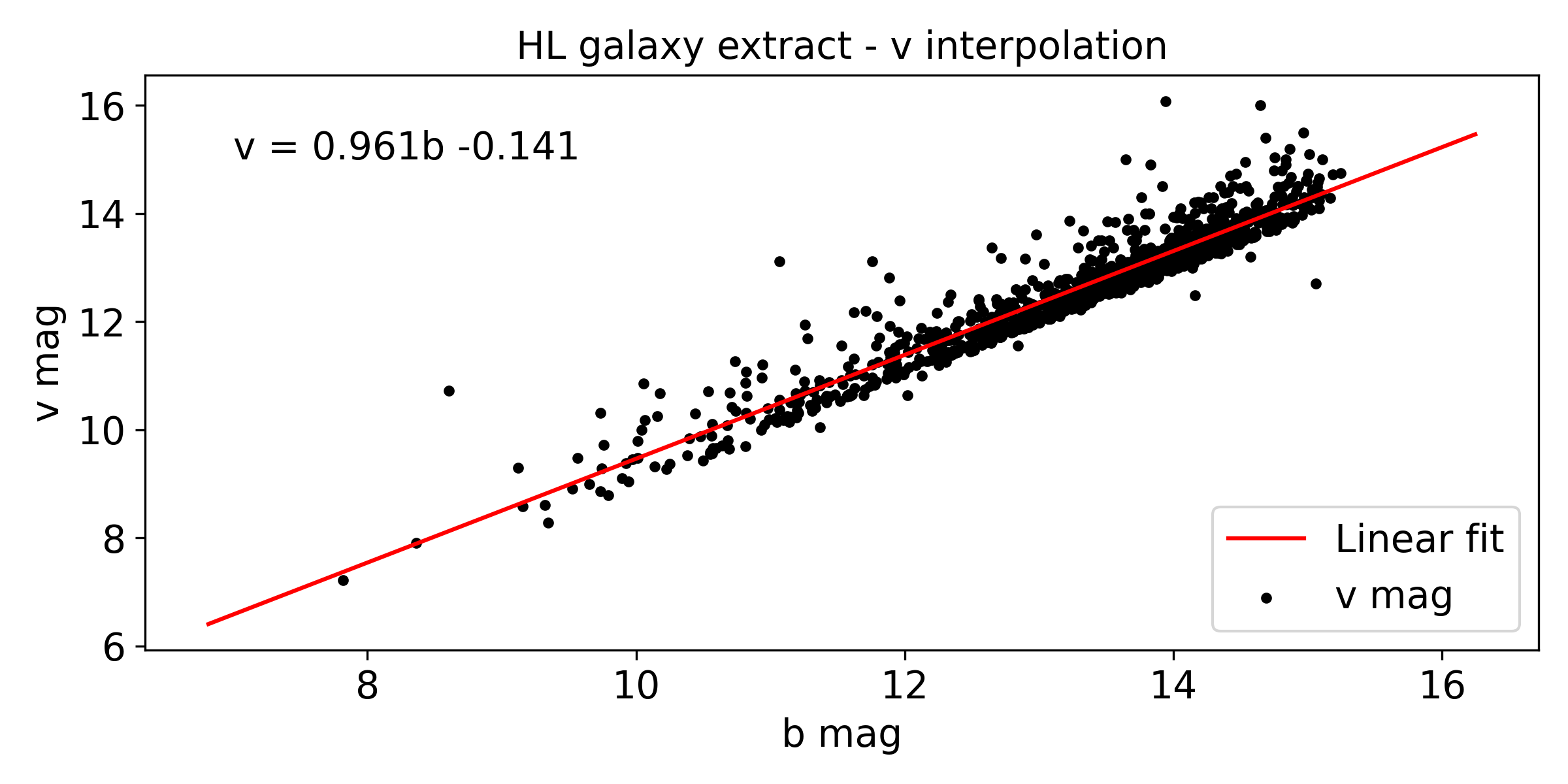}
    \caption{Hyperleda galaxy extract, apparent V mag vs B mag. $\sim$25\% of entries had existing measured values, so a simple linear interpolation from these was used to estimate absent V mag values.}
    \label{fig:HGE_appV_appB}
\end{figure}

As some of these galaxies lacked measured V band magnitudes, a linear relationship between $m_\text{B}$ and $m_\text{V}$, based on those galaxies with both, was estimated (see \cref{fig:HGE_appV_appB}). 

\begin{figure}
    \centering
    \includegraphics[width=0.95\linewidth]{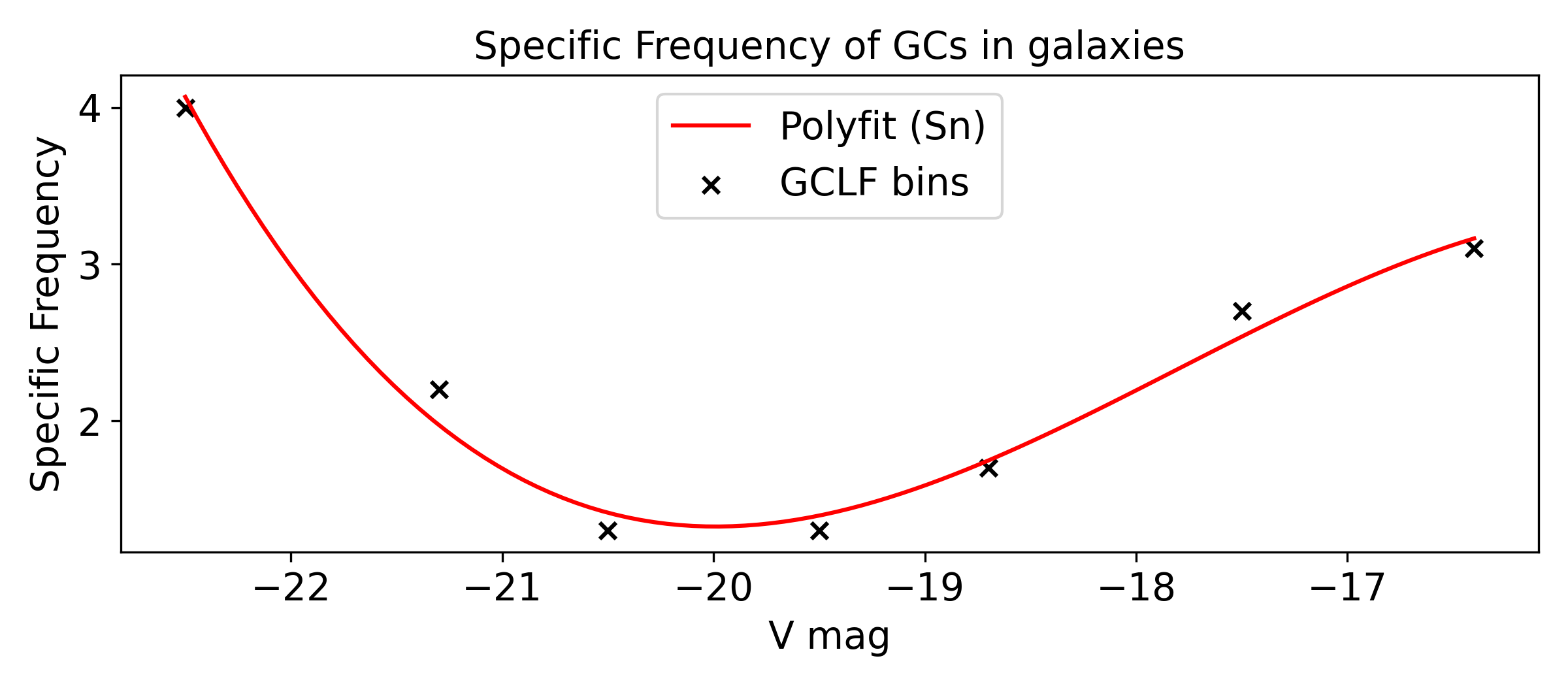}
    \caption{Specific frequency, $S_\text{N}$, defined by \citet{Harris_1981}, of globular clusters from bins measured by \citet{Peng_2008}. $S_\text{N}$ is numbers of clusters per unit galaxy luminosity. A polynomial fit was made to estimate $S_\text{N}$ for our galaxy sample.}
    \label{fig:SF_Peng}
\end{figure}

Using these values with the Hyperleda distance modulus, the absolute $M_\text{V}$ was used to determine the specific frequency, $S_\text{N}$, of the GC/CSS population using the trend determined by \citet[Table 3,][]{Peng_2008} from sample bins for their Virgo cluster (ACSVCS) galaxies. (see \cref{fig:SF_Peng}). 
Finally, using $S_\text{N}$, the relation detailed in \citet{Harris_1981} between GC population and galaxy V-band luminosity 
was rearranged 
to estimate the number of GCs hosted by each galaxy in our sample, giving
\begin{equation}\label{GC_per_host_estimate}
    N_\text{GC}=\frac{S_\text{N}}{10^{0.4(M_\text{V}+15)}}
\end{equation}
and from this the total number of GCs hosted by the 3714 galaxies in the sample was estimated as
\begin{equation}\label{GC_total_estimate}
    N_\text{GC,Total}=\sum\frac{S_\text{N}}{10^{0.4(M_\text{V}+15)}}\approx2.8\times10^6
\end{equation}



We then estimated the number of GCs above a mass of $10^6\Msun$ for each galaxy, following the process based on \citet[Fig.9 \& 11, \& Eq.18.]{Jordan_2007}. 
It is observed that the turnover magnitude of the approximately gaussian GCLF is only weakly dependent on the host galaxy mass \citep{Jordan_2007}. Hence, given the limited galaxy mass range sampled here we use a fixed value of $\mu_g = -7.2$ mag. In contrast, the dispersion of the GCLF is observed to vary sufficiently that we calculate an individual value for each galaxy following the relation $\sigma_g = 1.14 - (0.1\times (M_\text{B,gal}+20))$ from \citet{Jordan_2007}.

With these normal distribution parameters, for each host galaxy, we calculated the average number of GCs hosted which would be above a mass of $10^6\Msun$. For this we used a g-band mass-to-light ratio of 3.25 from \citet{Maraston_1998, Maraston_2005}, in respect of an old (10 Gyr), metal-poor ([Z/H] = -1.35) population based on a Salpeter IMF. We estimate the number of GCs above a mass of $10^6\Msun$ as $N_\text{gc,>1\text{E}6} \approx \num{2.4E5}$.


Consolidated light curves for transient alert objects were taken from the public alert archives from the period of the first phase of the ZTF programme (ZTF-1), lasting 30 months from Mar 2018 to Sep 2020, with additional data from Oct 2020 to Nov 2023. The ZTF facility aims to provide a full northern-sky survey cadence of 3-days, however, to account for inevitable gaps in observational coverage, the alert streams for potential GCs had a correction applied based on observations of the host galaxy.

A recent publication by \citet{Bandopadhyay_2024} details simulations which suggest the time to peak for TDEs is determined ``almost exclusively'' by the mass of the BH. They further suggest the time to peak is proportional to the square root of \MBH, with full disruption of larger stars being the most luminous, but partial disruptions potentially lasting `longer'. Their estimate for $\MBH = 10^6 \Msun$ is 30 days to peak and would give a peak time for $\MBH=10^5 \Msun$ of 9 days, and $\MBH=10^4 \Msun$ of 3 days. The fall-time of the curve is normally seen to follow roughly -5/3 power law decay, but can be longer for partial disruptions \citep{Mummery_2024}, so if we assume the fall-time is a few times the time to peak, we should observe $\MBH>10^4 \Msun$, although $\MBH<10^4 \Msun$ may be marginal, and difficult to classify. Based on this, a threshold of 9 days or more (i.e., 3 $\times$ 3 day nominal cadence) was defined as a ZTF `gap'. This was judged to be an extremely conservative estimate of the minimum amount of time a TDE around an IMBH should remain observable given the ZTF magnitude limit. The sum of gaps for each host target was deducted from the total observation period used in the final TDE rate estimation (see \cref{discussion}).

\section{Transient Alert analysis}
\label{Alert Analysis}
Using the ZTF alert archive we searched around the host galaxy positions to identify isolated alerts outside the galaxy nucleus, and also to check for misclassified TDEs in the search area. TDEs involving SMBH have been observed to peak at $-22\lesssim\!M_V\lesssim-17$ \citep[e.g.][]{vanVelzen_2020, Hammerstein_2023}. This peak is attained over a period of \textasciitilde1 month, and the TDE remains visible for many months. At the outer limit of our survey volume (<120 Mpc) this gives a peak of $\simeq\!16$ mag. Consequently, in the absence of internal dust, given the ZTF median sensitivity of 20.8 mag in g-band, 20.6 mag in r-band \citep{Bellm_2019a} we expect any SMBH TDEs in the search region to be observable. 
As noted by \citet{Metzger_2016}, there is a weak dependence of peak luminosity on \MBH, but adiabatic losses from radiation passing through the outer envelope of ejecta in lower mass BH interactions may suppress the peak optical luminosity. This is due to the model where post-peak TDE light curves follow the matter fallback rate onto the BH, although there is some debate over this $\MBH$ correlation \citep[e.g.][]{Nicholl_2020}. Nevertheless, we assume a nominal 2-3 mag reduction in apparent magnitude compared to current SMBH observations, and given the cadence and magnitude limit of the ZTF, a minimum observability period of \textasciitilde9 days for IMBH related TDEs was not considered unreasonable \citep[e.g.][]{Angus_2022}. 

\begin{figure}
    \centering
    \includegraphics[width=1\linewidth]{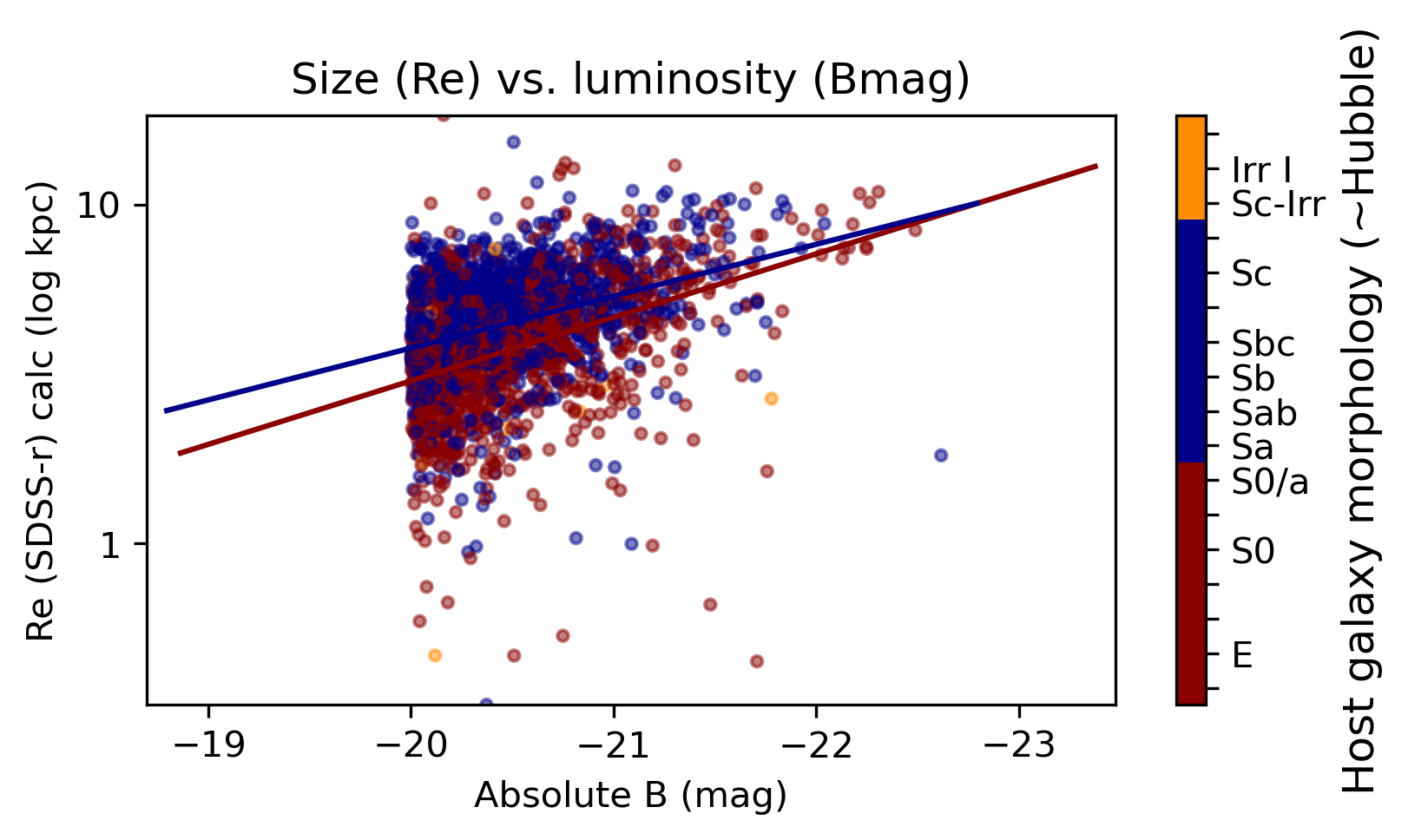}
    \caption{Size ($R_e$) vs. Bmag plot used to estimate the effective radius of host galaxies, for which SDSS petrosian $R_\text{50}$ and $R_\text{90}$ magnitudes were not available. }
    \label{fig:Gal_sizeVsLum_wtype}
\end{figure}

SDSS parameters were downloaded for each of the host galaxies where available. Following the prescription determined by \citet{Graham_2005}, the effective radius $R_e$ of the galaxy was determined in relation to the SDSS petrosian magnitudes $R_{50}$ and $R_{90}$ through
\begin{equation}
    R_e \approx \frac{R_{50}}{1 - P_3 (R_{90} / R_{50})^{P_4}}
\end{equation}
where P3 and P4 equal $8.0\times10^{-6}$ and $8.47$, respectively. The SDSS coverage was 2602 entries of the 3714 original host galaxies in the list, and so a linear fit to the B-mag of the host was estimated from the available $R_e$ values following a similar logic to that outlined by \citet{Shen_2003}, to calibrate the $R_e$ vs. B-mag relation (\cref{fig:Gal_sizeVsLum_wtype}). A split on host galaxy morphology was included at type code=0 (S0/a).

Although the extent of a GC system relative to its host galaxy varies between galaxy type and mass, many authors have observed and assumed values in the range $7-10R_e$ \citep[e.g.][]{Faifer_2011,Forbes_2017, Bilek_2019} and thus a value of $10R_e$ was assumed here. This value was used to ensure coverage of all GCs associated with a host galaxy by estimating the galaxy extent and ultimately set the search cone radius for alerts.



The cone search was applied against the ALeRCE (Automatic Learning for Rapid Classification of Events) host site using API scripts \citep{Forster_2021}. Alerts with ZTF real/bogus rating $rb\!<\!0.5$ were rejected, and remaining alert data downloaded for further analysis. For all alerts within a galaxy region, ZTF metadata was used to aid transient candidate prioritising. This included the number of alerts within a region, a count of unique ZTF transient objects associated with the alerts and the maximum detections associated with a single object. The latter was used to reject objects where the detection history was less than 3, \ie a minimum of 3 transient alerts were required, and single detections were ignored. Alerts in proximity ($<\ang{;;2}$) of solar system objects were also rejected. 


Final filtering was achieved by rejecting ZTF alert objects within $\ang{;;2}$ of GAIA DR3 objects with a measured parallax (4741), on the assumption these objects are `local'. Additionally, ZTF objects previously classified in TNS (289) were only given a cursory inspection.

The volume around galaxy peripheries furnished alerts with various light curves, including a large number of supernova (SN) (see \cref{tab:tns_summary}). The alert objects were explored by cross-matching with external databases, where additional data extracted included Transient Name Server (\href{https://www.wis-tns.org}{TNS}) information, and correlation with \href{http://simbad.cds.unistra.fr/simbad/}{SIMBAD} \citep{Wenger_2000} and GAIA databases. 

\begin{figure*}
    \centering
    \includegraphics[width=0.78\textwidth]{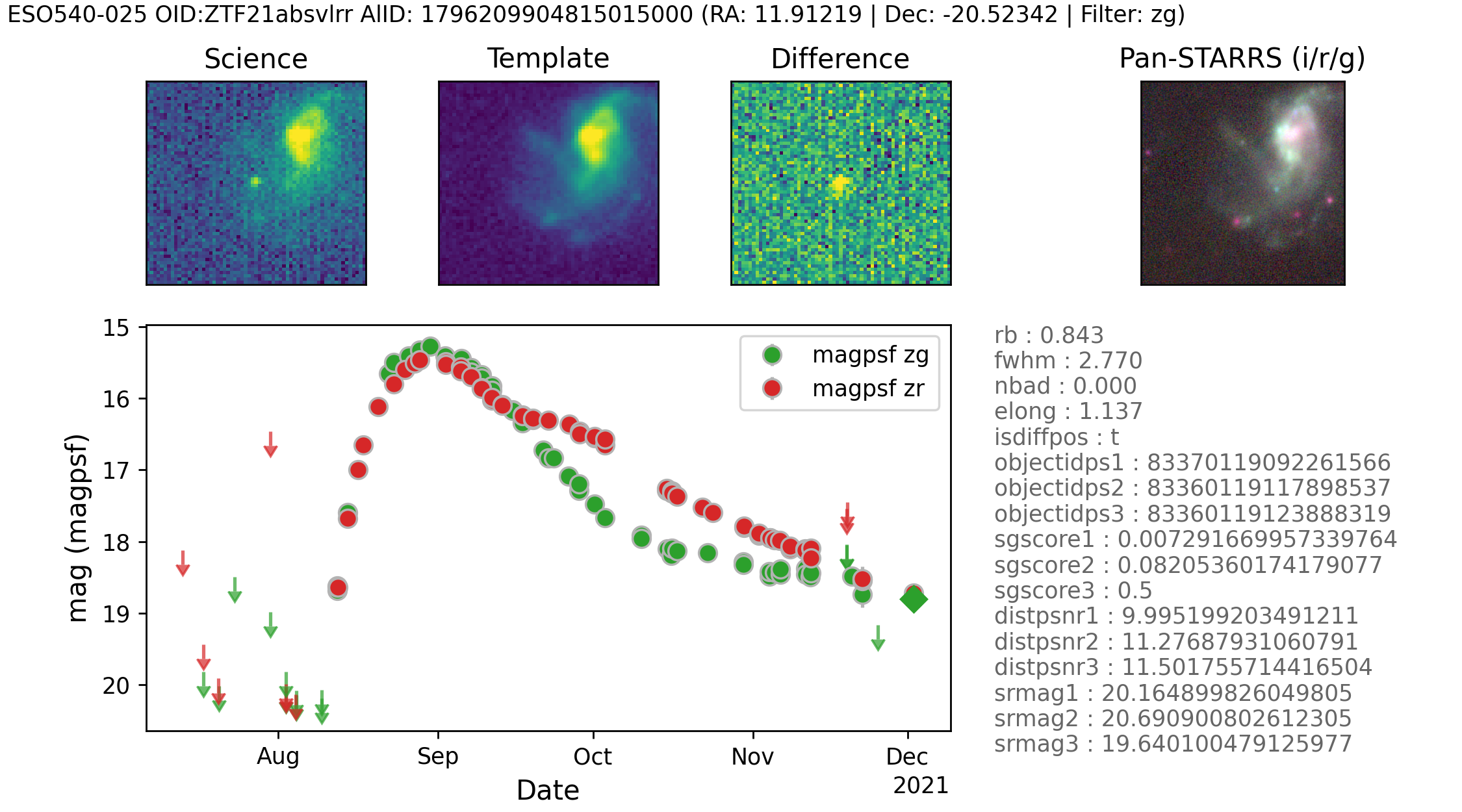}
    \caption[Visualisation of ZTF alert data, shown for ZTF21absvlrr within ESO540-025]{Visualisation of ZTF alert data, shown for ZTF21absvlrr within ESO540-025. 1' cutouts for science, reference and difference images are shown alongside PanSTARRS colour (gri) image. ID for latest alert ID (AlID) given and ZTF alert metadata included. This candidate was classified in the TNS records as a Type 1a SN.}
    \label{fig:ESO540-025_ztfAlerts}
\end{figure*}

The environment and lightcurves of all ZTF alert objects (1900) in host galaxies associated with the remaining alerts (65035) were inspected for evidence of TDEs. This was achieved using ZTF alert data in conjunction with ALeRCE ZTF Explorer (see \url{https://alerce.online/}). An example of ZTF alert data is shown in \cref{fig:ESO540-025_ztfAlerts} for galaxy ESO540-025, a Type 1a SN. The figure shows visualization of ZTF alert history with $\ang{;1;}$ final alert image cutouts and PS combined $gri_\text{PS1}$ image, all centered on the alert position. 
. 

The process followed to classify the transients associated with the ZTF alert objects, involved inspection of all of the 1900 non-Gaia parallax lightcurves. An initial pass of the lightcurves discounted those based on their proximity to the nucleus of the host galaxy (785). A second criteria applied at this time was to discount spurious transients (990), defined as objects which nominally showed no obvious signs of a curve, had minimal total ($<4$) real detections or contained multiple ($>2$) intermediate non-detections between real detections.

For the remaining 125 ZTF alert objects, further scrutiny was applied. Each of these were placed in bins with the following precedence:
\begin{itemize}
    \item Artefact: These are clear imaging artefacts in either the science or template images, for example, even after masking caused by diffraction spikes from a nearby bright star.
    \item Star: Stellar objects (some variable) picked up in field. Confirmed with Gaia proximity and source extraction assessment.
    \item Background galaxy: Galaxies confirmed in background to host galaxy as viewed in PanSTARRS imagery and through source extraction assessment.
    \item Positive Template: Some ZTF objects were active supernova events while reference templates were being generated early in 2018. Subsequent science imagery consists of dimmer magnitude values, and hence constant negative difference alerts. These were scrutinised but ignored.
    \item Spiral / Halo: Transients located predominantly in spiral arms, often star forming regions or late-type galaxies or bulges of early type galaxies, in addition to some halo transients showing no distinct curve. A minimal number (<10) of the objects in this category showed distinct light curves and were given further scrutiny, validating the rise/fall time  (typically longer than SNe), colour ($-0.4\lesssim g-r\lesssim0$) and colour evolution (which for TDEs is practically constant) of the object \citep{vanVelzen_2021}, in addition to the location relative to the host galaxy. All were rejected.
    \item Indeterminate: Remaining (miscellaneous) bin of light curves, defying classification under other criteria, but with no definitive shape to alerts associated with ZTF object, often with multiple intermediate non-detections.
\end{itemize}
Further details and examples of the above classifications are included in \cref{Light curve categorisation}.
 
\begin{table} 
\centering
\caption[Summary of TNS classifications]{Summary of ZTF alert object classifications.}
{
\small
\begin{tabular}{@{}ccp{5.7cm}@{}}
Count & Sub count & Description \\ \midrule
2135 & & Post-filter galaxies with ZTF alert objects \\
6913 & & Distinct ZTF alert objects\\ \midrule
4741 & & Objects with Gaia parallax \\
289 & & TNS classified entries \\ \midrule
& 161 & TNS SN type II / pec / P / b / n / 1b / 1c \\ 
& 122 & TNS SN type 1a \\
& 3 & Other TNS (ILRT / gap / LRN) \\
& 1 & Nova \\
& 2 & Tidal Disruption Event (Nuclear) \\ \midrule
1900 & & TNS unclassified transients \\ \midrule
 & 990 & Spurious (minimal transients, no distinct curve) \\
 & 785 & Nuclear proximity (bulge \& nuclear transients) \\
 & 58 & Spiral arm / Halo location \\
 & 21 & Background galaxy AGN \\
 & 14 & Star / Galaxy artefacts \\
 & 13 & Indeterminate (no curve) \\
 & 11 & Star (Variable) \\
 & 8 & +ve template (SN occurred in ZTF ref. image)
\end{tabular}
}
\label{tab:tns_summary}
\end{table}

No halo TDEs were evident in our search of alerts proximal to filtered galaxies out to 120 Mpc, although 2 TNS classified nuclear TDEs were noted (ZTF18aabkvpi and ZTF20aamaczu in NGC3800). At the date of this report, a search of the TNS database also showed 77 transients classified as TDEs, the discovery of 46 of which were credited to ZTF data. The light curve from the closest TDE  \citep[AT2019qiz, $\sim$66 Mpc,][]{Forster_2019, Nicholl_2020, Hung_2021} is an example associated with an SMBH (only 1 dex outside the IMBH mass range). There are many variables associated with both BH and disrupted star mass, and impact parameters of the encounter, leading to a range of outcomes and profile evolution for the emissions from the event \citep[e.g.][]{Ulmer_1999,Guillochon_2013,Malyali_2019}. Nevertheless, to a first order we can argue that TDEs present in the region searched would be detected within this data, However, as noted in \cref{sample}, if the simulations of \citet{Bandopadhyay_2024} are correct, then with our conservative observability limits (\ie >9-day threshold defined as an gap in observations), detection of TDEs associated with IMBH$<10^4$ would be marginal.

\section{Discussion}\label{discussion}
Early predictions of TDE rates, albeit for SMBH and solar mass stars, were quoted at $n_\text{TDE}\!\approx\!10^{-4} \text{gal}^{-1}\text{yr}^{-1}$, e.g. \citet{Frank_1976, Ulmer_1999}. 
\citet{Stone_2016} note observational results suggest an order of magnitude lower rate at $n_\text{TDE}\approx\!10^{-5} \text{gal}^{-1}\text{yr}^{-1}$. Recent observations using ZTF data have provided some validation to this lower rate, with an average optical rate of $\approx\!3.2\times10^{-5} \text{gal}^{-1}\text{yr}^{-1}$ determined by \citet{Yao_2023}. Conversely, as the rate is sensitive to the central density profile of a galaxy, \citet{Wang_2004} suggested for systems such as nucleated dwarfs, with steep nuclear density profiles, the rate may be higher than predicted by \citet{Frank_1976} at $n_\text{TDE}\approx\!10^{-3} \text{gal}^{-1}\text{yr}^{-1}$. This latter case is particularly relevant for the composite population of CSS, thus we compare the above rates with an upper limit based on our non-detection of TDEs. 

The period of data sampling, $t_\text{ZTF}$, for the ZTF alert pipeline was taken to be mid-March 2018 to mid-Nov 2023 or 68 months (5.7 yrs), however, as detailed in \cref{sample}, sample periods of GC population observations were adjusted for host galaxy observability. Using these adjusted timescales, and non-detection of TDEs in the total notional clusters of $N_\text{GC,Total}\approx\!2.8\times10^6$, we estimate a total TDE upper limit of 
\begin{equation}\label{GC_vis_TDE}
    N_\text{TDE,Total}<\frac{1}{N_\text{GC,Total}\times t_\text{ZTF}}\lessapprox2\times10^{-7} \text{CSS}^{-1}\text{yr}^{-1}
\end{equation}
which is \textasciitilde 2 dex below the lowest literature detection limit for galaxies. This implies either \textasciitilde 100 TDEs in this sample occurred but were undetected, or the TDE rate in CSS's is considerably lower than in galaxies.

Despite attempting to ensure object visibility, it is possible non-detections may occur if the immediate vicinity of the BH is optically thick \citep[i.e.][]{Panagiotou_2023}. However, a surplus of gas and dust is at odds with the view of GCs, 
as excess material is considered to be used or ejected from these systems during their primary star formation episode. Conversely, for CSS formed through tidal stripping, more gas and dust may temporarily be present in the core due to inflows after interaction of the progenitor with a larger galaxy \citep[e.g.][]{Janz_2016,Du_2019}, although in general, dust is not expected to be significant. No significant amount has been conclusively detected in any objects of GC/UCD mass. It does not survive in GCs/UCDs, being pushed out by radiation pressure rapidly \citep[e.g. timescales of $\sim\!10^5$ yr,][]{McDonald_2009}, the same is true of a stripped CSS once it has lost its dark matter halo and the potential becomes that provided by the stars alone. It is possible that a thick disc, would be sufficient to absorb optical emissions, although statistically, face on viewing should have given rise to detections \citep[see][]{Dai_2018}. 


In the case where the TDE rate is lower in CSS's than galaxies, there are two possible explanations; IMBH do not exist in CSS, or IMBH do exist, but the TDE rate is suppressed in this environment by some unknown mechanism.

When considering whether IMBH exist in CSS, it is useful to split the composite population into massive clusters and tidally stripped galaxies. It has long been thought GCs (and thus massive clusters) may harbour nuclear BH and at the lower end of the CSS mass scale this view has tantalising implications. If central BHs form in clusters following a similar process to larger galaxies, we can extrapolate BH mass-scaling relationships to intermediate scales. There is, however, a lack of conclusive evidence of nuclear BH in clusters. Early studies suggested formation of IMBH in clusters was possible through core collapse and runaway mergers of massive stars \citep{PortegiesZwart_2002}, or repeated merging of mass segregated BHs \citep{Miller_2002}. Nevertheless, numerical simulations \citep[e.g.][]{Giesler_2018} show dynamical interactions in cluster cores would eject many BHs before larger mass BH could `grow' through mergers. X-ray emissions also seem prevalent in GCs \citep[e.g.][]{Strader_2012a}, most associated with compact object related low-mass X-ray binaries (LMXB). Thus, an excess of dynamical mass compared to luminous mass, might be explained by a population of core segregated stellar mass BH and neutron stars rather than a single nuclear IMBH. Several examples exist of observational evidence fitting both scenarios \citep[e.g.][]{Vitral_2021}. 

There is much evidence to suggest the TDE rate in galaxies appears dominated by galaxies with a high surface mass density. Authors such as, \citet[Table 3]{Graur_2018} find TDE host properties $\approx\!10^9\!\leq\!\Sigma_{M_\star}\!\leq\!10^{10}\Msun\ \text{kpc}^{-2}$. \citet{Law-Smith_2017} discuss the stellar mass density in relation to the Sérsic index as a broad indicator of the galaxy’s light profile, and details in \citet{French_2020} suggest a range of $10^{10.5}-10^{11}\Msun\ \text{kpc}^{-2}$, although their findings are based on a detailed HST analysis of only 4 TDE hosts.  

In comparison, we also find there is much evidence that the high mass end of GCs / UCDs, \ie those above a mass of $10^6\Msun$, have surface mass densities comparable to galaxies, especially those formed through tidal stripping processes. For example, \citet[Figure 14]{Norris_2014} find for GCs/UCDs$>10^6\Msun$ the surface mass density within $R_e\approx 10^9\!\leq\!\Sigma_{M_\star}\!\leq\!10^{11}\Msun\ \text{kpc}^{-2}$. Additionally, in UCDs where detailed observations and analysis has suggested the presence of a black hole, noted in \cref{intro}, the stellar density in the immediate proximity of the BH are shown to be high. For example, \citet{Afanasiev_2018} find Fornax-UCD3 has a stellar mass density $\approx\!2.5\times\!10^{10}\Msun\ \text{kpc}^{-2}$ within a pc of the centre. \citet{Ahn_2018} find an order of magnitude greater surface mass density for M59-UCD3, and similarly for VUCD3 \citep{Ahn_2018}. These comparisons support the assumption that the TDE rate in GCs/UCDs $>10^6\Msun$ will be comparable to the ‘normal’ rate in high surface mass density galaxies.

At the upper end of the CSS mass scale, considering UCDs or cEs formed through tidal stripping or cEs formed as low-mass galaxies, larger mass BHs are anticipated, and have been detected, as mentioned in \cref{intro}. Bulges and cores surviving tidal threshing would have a nuclear BH mass related to the progenitor, and thus higher than expected for the remnant cluster mass. This is a differentiator of formation in the composite CSS population, and an example of core phenomenological parameters which are considered to remain unaffected by tidal stripping. Therefore, given the expected BH masses and high surface mass density, there is no reason to expect fewer TDEs in this type of environment than the galactic cores for which the theoretical rate has been measured. Authors such as \citet{Janz_2015} have identified UCDs with increased dynamical to stellar mass ratios with nuclear SMBH, and if these have the same TDE rate as ‘normal’ galaxies our non-detection, allows us to set a limit on the number of stripped UCDs in the area searched. 

In \cref{sample}, we used the GCLF to estimate the number of GCs in the volume searched above a mass of $10^6\Msun$, to be $N_\text{gc,>1\text{E}6} \approx \num{2.4E5}$. Further to this, we then use an average optical TDE rate of $\approx\num{3.2E-5}\text{gal}^{-1}\text{yr}^{-1}$ determined by \citet{Yao_2023}, and also assume a conservative $40\%$ ZTF observability figure over the 5.7 years for the brighter and longer period TDEs involving SMBH in stripped UCDs. From this we set an upper limit on the number of stripped UCDs populating the galaxies in the area searched as 
\begin{align}
    N_\text{GC, Strip} \lessapprox \frac{1}{\num{3.2E-5} \times 5.7 \times 0.4} \lessapprox \num{1.4E4}.
\end{align}

We estimate this represents an upper limit for the fraction of UCDs, as a proportion of the population of GCs over $10^6 \Msun$, to be
\begin{align}
    \varphi_\text{GC,strip} \lessapprox \num{1.4E4} / \num{2.4E5} \lessapprox 5.8\%
\end{align}
 
This is in reasonable agreement with the semi-analytic model and observations of \citet{Pfeffer_2016}, who suggested stripped nuclei UCDs account for 2.5\% of the GC / UCD total between $10^6-10^7\Msun$ increasing to 40\% above $10^7 \Msun$.
 

The approach utilised here of searching for TDEs in CSS will naturally lead to detections or increasingly stringent upper limits as the duration of ZTF observations increases. In addition, the advent of the Vera Rubin Observatory (LSST) will dramatically increase the survey volume probed thanks to it's greater depth (\textasciitilde 24 mag)\citep{Ivezic_2019} and approximately halved seeing (0.75 vs 1.4). 


\section{Conclusions}
In this paper we have searched for optical transients consistent with being TDEs associated with IMBH/SMBHs in CSSs using 68 months of ZTF archive data. The search area consisted of the region within 10R$_e$ of massive galaxies ($M_\text{B}\! <\! -20\ \text{mag}$) at distance $2\! \leq\! D\! \leq\! 120$ Mpc, notionally to encompass the entire population CSSs hosted by the galaxies.  No transient alert profiles were found to be consistent with being non-nuclear TDEs and hence we conclude an upper limit for the rate of TDEs in CSS is $n_\text{TDE,Total}\lessapprox\!10^{-7} \text{CSS}^{-1}\text{yr}^{-1}$. We note this rate is two dex below the observed TDE rate involving SMBH interacting with solar mass main sequence stars in the nucleus of galaxies. We also estimated an upper limit on the number of UCDs formed through tidal stripping processes to be $N_\text{GC, Strip} < \num{1.4E4}$ which we determined represents an upper limit for the fraction of tidally stripped type UCDs as a proportion of the population of CSSs over $10^6 \Msun$ as $\varphi_\text{GC,strip} < 5.8\%$

\section*{Acknowledgements}

We thank the anonymous referee for their helpful and insightful comments.

Based on observations obtained with the Samuel Oschin 48-inch Telescope at the Palomar Observatory as part of the Zwicky Transient Facility project. ZTF is supported by the National Science Foundation under Grant No. AST-1440341 and a collaboration including Caltech, IPAC, the Weizmann Institute for Science, the Oskar Klein Center at Stockholm University, the University of Maryland, the University of Washington, Deutsches Elektronen-Synchrotron and Humboldt University, Los Alamos National Laboratories, the TANGO Consortium of Taiwan, the University of Wisconsin at Milwaukee, and Lawrence Berkeley National Laboratories. Operations are conducted by COO, IPAC, and UW.

We acknowledge the usage of the HyperLeda database \url{http://leda.univ-lyon1.fr/} \citep{Makarov_2014}.

We acknowledge the use of data from the ALeRCE project \url{https://alerce.science}. 

We acknowledge the use of data from the Sloan Digital Sky Survey (SDSS) \url{http://www.sdss.org/}. Funding for the Sloan Digital Sky Survey (SDSS) has been provided by the Alfred P. Sloan Foundation, the Participating Institutions, the National Aeronautics and Space Administration, the National Science Foundation, the U.S. Department of Energy, the Japanese Monbukagakusho, and the Max Planck Society.

\section*{Data Availability}
The data used in this analysis is available at  \url{https://github.com/Pommers/IMBH}.



\bibliographystyle{mnras}
\bibliography{imbh_search} 

\begin{thebibliography}{}
\makeatletter
\relax
\def\mn@urlcharsother{\let\do\@makeother \do\$\do\&\do\#\do\^\do\_\do\%\do\~}
\def\mn@doi{\begingroup\mn@urlcharsother \@ifnextchar [ {\mn@doi@} {\mn@doi@[]}}
\def\mn@doi@[#1]#2{\def\@tempa{#1}\ifx\@tempa\@empty \href {http://dx.doi.org/#2} {doi:#2}\else \href {http://dx.doi.org/#2} {#1}\fi \endgroup}
\def\mn@eprint#1#2{\mn@eprint@#1:#2::\@nil}
\def\mn@eprint@arXiv#1{\href {http://arxiv.org/abs/#1} {{\tt arXiv:#1}}}
\def\mn@eprint@dblp#1{\href {http://dblp.uni-trier.de/rec/bibtex/#1.xml} {dblp:#1}}
\def\mn@eprint@#1:#2:#3:#4\@nil{\def\@tempa {#1}\def\@tempb {#2}\def\@tempc {#3}\ifx \@tempc \@empty \let \@tempc \@tempb \let \@tempb \@tempa \fi \ifx \@tempb \@empty \def\@tempb {arXiv}\fi \@ifundefined {mn@eprint@\@tempb}{\@tempb:\@tempc}{\expandafter \expandafter \csname mn@eprint@\@tempb\endcsname \expandafter{\@tempc}}}

\bibitem[\protect\citeauthoryear{{Abbott} et~al.,}{{Abbott} et~al.}{2020}]{Abbott_2020}
{Abbott} R.,  et~al., 2020, \mn@doi [\prl] {10.1103/PhysRevLett.125.101102}, \href {https://ui.adsabs.harvard.edu/abs/2020PhRvL.125j1102A} {125, 101102}

\bibitem[\protect\citeauthoryear{{Afanasiev} et~al.,}{{Afanasiev} et~al.}{2018}]{Afanasiev_2018}
{Afanasiev} A.~V.,  et~al., 2018, \mn@doi [\mnras] {10.1093/mnras/sty913}, \href {https://ui.adsabs.harvard.edu/abs/2018MNRAS.477.4856A} {477, 4856}

\bibitem[\protect\citeauthoryear{{Ahn} et~al.,}{{Ahn} et~al.}{2017}]{Ahn_2017}
{Ahn} C.~P.,  et~al., 2017, \mn@doi [\apj] {10.3847/1538-4357/aa6972}, \href {https://ui.adsabs.harvard.edu/abs/2017ApJ...839...72A} {839, 72}

\bibitem[\protect\citeauthoryear{{Ahn} et~al.,}{{Ahn} et~al.}{2018}]{Ahn_2018}
{Ahn} C.~P.,  et~al., 2018, \mn@doi [\apj] {10.3847/1538-4357/aabc57}, \href {https://ui.adsabs.harvard.edu/abs/2018ApJ...858..102A} {858, 102}

\bibitem[\protect\citeauthoryear{{Angus} et~al.,}{{Angus} et~al.}{2022}]{Angus_2022}
{Angus} C.~R.,  et~al., 2022, \mn@doi [Nature Astronomy] {10.1038/s41550-022-01811-y}, \href {https://ui.adsabs.harvard.edu/abs/2022NatAs...6.1452A} {6, 1452}

\bibitem[\protect\citeauthoryear{{Bade}, {Komossa}  \& {Dahlem}}{{Bade} et~al.}{1996}]{Bade_1996}
{Bade} N.,  {Komossa} S.,   {Dahlem} M.,  1996, \aap, \href {https://ui.adsabs.harvard.edu/abs/1996A&A...309L..35B} {309, L35}

\bibitem[\protect\citeauthoryear{{Bandopadhyay} et~al.,}{{Bandopadhyay} et~al.}{2024}]{Bandopadhyay_2024}
{Bandopadhyay} A.,  et~al., 2024, \mn@doi [\apjl] {10.3847/2041-8213/ad0388}, \href {https://ui.adsabs.harvard.edu/abs/2024ApJ...961L...2B} {961, L2}

\bibitem[\protect\citeauthoryear{{Bekki}, {Couch}  \& {Drinkwater}}{{Bekki} et~al.}{2001}]{Bekki_2001}
{Bekki} K.,  {Couch} W.~J.,   {Drinkwater} M.~J.,  2001, \mn@doi [\apjl] {10.1086/320339}, \href {https://ui.adsabs.harvard.edu/abs/2001ApJ...552L.105B} {552, L105}

\bibitem[\protect\citeauthoryear{{Bellm} et~al.,}{{Bellm} et~al.}{2019a}]{Bellm_2019b}
{Bellm} E.~C.,  et~al., 2019a, \mn@doi [\pasp] {10.1088/1538-3873/aaecbe}, \href {https://ui.adsabs.harvard.edu/abs/2019PASP..131a8002B} {131, 018002}

\bibitem[\protect\citeauthoryear{{Bellm} et~al.,}{{Bellm} et~al.}{2019b}]{Bellm_2019a}
{Bellm} E.~C.,  et~al., 2019b, \mn@doi [\pasp] {10.1088/1538-3873/ab0c2a}, \href {https://ui.adsabs.harvard.edu/abs/2019PASP..131f8003B} {131, 068003}

\bibitem[\protect\citeauthoryear{{B{\'\i}lek}, {Samurovi{\'c}}  \& {Renaud}}{{B{\'\i}lek} et~al.}{2019}]{Bilek_2019}
{B{\'\i}lek} M.,  {Samurovi{\'c}} S.,   {Renaud} F.,  2019, \mn@doi [\aap] {10.1051/0004-6361/201834675}, \href {https://ui.adsabs.harvard.edu/abs/2019A&A...625A..32B} {625, A32}

\bibitem[\protect\citeauthoryear{{Bloom} et~al.,}{{Bloom} et~al.}{2011}]{Bloom_2011}
{Bloom} J.~S.,  et~al., 2011, \mn@doi [Science] {10.1126/science.1207150}, \href {https://ui.adsabs.harvard.edu/abs/2011Sci...333..203B} {333, 203}

\bibitem[\protect\citeauthoryear{{Chambers} et~al.,}{{Chambers} et~al.}{2016}]{Chambers_2016}
{Chambers} K.~C.,  et~al., 2016, arXiv e-prints, \href {https://ui.adsabs.harvard.edu/abs/2016arXiv161205560C} {p. arXiv:1612.05560}

\bibitem[\protect\citeauthoryear{{Dai}, {McKinney}, {Roth}, {Ramirez-Ruiz}  \& {Miller}}{{Dai} et~al.}{2018}]{Dai_2018}
{Dai} L.,  {McKinney} J.~C.,  {Roth} N.,  {Ramirez-Ruiz} E.,   {Miller} M.~C.,  2018, \mn@doi [\apjl] {10.3847/2041-8213/aab429}, \href {https://ui.adsabs.harvard.edu/abs/2018ApJ...859L..20D} {859, L20}

\bibitem[\protect\citeauthoryear{{Di Carlo} et~al.,}{{Di Carlo} et~al.}{2021}]{DiCarlo_2021}
{Di Carlo} U.~N.,  et~al., 2021, \mn@doi [\mnras] {10.1093/mnras/stab2390}, \href {https://ui.adsabs.harvard.edu/abs/2021MNRAS.507.5132D} {507, 5132}

\bibitem[\protect\citeauthoryear{{Du} et~al.,}{{Du} et~al.}{2019}]{Du_2019}
{Du} M.,  et~al., 2019, \mn@doi [\apj] {10.3847/1538-4357/ab0e0c}, \href {https://ui.adsabs.harvard.edu/abs/2019ApJ...875...58D} {875, 58}

\bibitem[\protect\citeauthoryear{{Evans} \& {Kochanek}}{{Evans} \& {Kochanek}}{1989}]{Evans_1989}
{Evans} C.~R.,  {Kochanek} C.~S.,  1989, \mn@doi [\apjl] {10.1086/185567}, \href {https://ui.adsabs.harvard.edu/abs/1989ApJ...346L..13E} {346, L13}

\bibitem[\protect\citeauthoryear{{Faifer} et~al.,}{{Faifer} et~al.}{2011}]{Faifer_2011}
{Faifer} F.~R.,  et~al., 2011, \mn@doi [\mnras] {10.1111/j.1365-2966.2011.19018.x}, \href {https://ui.adsabs.harvard.edu/abs/2011MNRAS.416..155F} {416, 155}

\bibitem[\protect\citeauthoryear{{Forbes}}{{Forbes}}{2017}]{Forbes_2017}
{Forbes} D.~A.,  2017, \mn@doi [\mnras] {10.1093/mnrasl/slx148}, \href {https://ui.adsabs.harvard.edu/abs/2017MNRAS.472L.104F} {472, L104}

\bibitem[\protect\citeauthoryear{{Forster}}{{Forster}}{2019}]{Forster_2019}
{Forster} F.,  2019, Transient Name Server Discovery Report, \href {https://ui.adsabs.harvard.edu/abs/2019TNSTR1857....1F} {2019-1857, 1}

\bibitem[\protect\citeauthoryear{{F{\"o}rster} et~al.,}{{F{\"o}rster} et~al.}{2021}]{Forster_2021}
{F{\"o}rster} F.,  et~al., 2021, \mn@doi [\aj] {10.3847/1538-3881/abe9bc}, \href {https://ui.adsabs.harvard.edu/abs/2021AJ....161..242F} {161, 242}

\bibitem[\protect\citeauthoryear{{Frank} \& {Rees}}{{Frank} \& {Rees}}{1976}]{Frank_1976}
{Frank} J.,  {Rees} M.~J.,  1976, \mn@doi [\mnras] {10.1093/mnras/176.3.633}, \href {https://ui.adsabs.harvard.edu/abs/1976MNRAS.176..633F} {176, 633}

\bibitem[\protect\citeauthoryear{{French}, {Arcavi}, {Zabludoff}, {Stone}, {Hiramatsu}, {van Velzen}, {McCully}  \& {Jiang}}{{French} et~al.}{2020}]{French_2020}
{French} K.~D.,  {Arcavi} I.,  {Zabludoff} A.~I.,  {Stone} N.,  {Hiramatsu} D.,  {van Velzen} S.,  {McCully} C.,   {Jiang} N.,  2020, \mn@doi [\apj] {10.3847/1538-4357/ab7450}, \href {https://ui.adsabs.harvard.edu/abs/2020ApJ...891...93F} {891, 93}

\bibitem[\protect\citeauthoryear{{Frieman} et~al.,}{{Frieman} et~al.}{2008}]{Frieman_2008}
{Frieman} J.~A.,  et~al., 2008, \mn@doi [\aj] {10.1088/0004-6256/135/1/338}, \href {https://ui.adsabs.harvard.edu/abs/2008AJ....135..338F} {135, 338}

\bibitem[\protect\citeauthoryear{{Gaia Collab.} et~al.,}{{Gaia Collab.} et~al.}{2016}]{GaiaCollaboration_2016}
{Gaia Collab.} et~al., 2016, \mn@doi [\aap] {10.1051/0004-6361/201629272}, \href {https://ui.adsabs.harvard.edu/abs/2016A&A...595A...1G} {595, A1}

\bibitem[\protect\citeauthoryear{{Gezari}}{{Gezari}}{2021}]{Gezari_2021}
{Gezari} S.,  2021, \mn@doi [\araa] {10.1146/annurev-astro-111720-030029}, \href {https://ui.adsabs.harvard.edu/abs/2021ARA&A..59...21G} {59}

\bibitem[\protect\citeauthoryear{{Gezari} et~al.,}{{Gezari} et~al.}{2006}]{Gezari_2006}
{Gezari} S.,  et~al., 2006, \mn@doi [\apjl] {10.1086/509918}, \href {https://ui.adsabs.harvard.edu/abs/2006ApJ...653L..25G} {653, L25}

\bibitem[\protect\citeauthoryear{{Giesler}, {Clausen}  \& {Ott}}{{Giesler} et~al.}{2018}]{Giesler_2018}
{Giesler} M.,  {Clausen} D.,   {Ott} C.~D.,  2018, \mn@doi [\mnras] {10.1093/mnras/sty659}, \href {https://ui.adsabs.harvard.edu/abs/2018MNRAS.477.1853G} {477, 1853}

\bibitem[\protect\citeauthoryear{{Gomez} \& {Gezari}}{{Gomez} \& {Gezari}}{2023}]{Gomez_2023}
{Gomez} S.,  {Gezari} S.,  2023, \mn@doi [\apj] {10.3847/1538-4357/acefbc}, \href {https://ui.adsabs.harvard.edu/abs/2023ApJ...955...46G} {955, 46}

\bibitem[\protect\citeauthoryear{{Gonz{\'a}lez}, {Kremer}, {Chatterjee}, {Fragione}, {Rodriguez}, {Weatherford}, {Ye}  \& {Rasio}}{{Gonz{\'a}lez} et~al.}{2021}]{Gonzalez_2021}
{Gonz{\'a}lez} E.,  {Kremer} K.,  {Chatterjee} S.,  {Fragione} G.,  {Rodriguez} C.~L.,  {Weatherford} N.~C.,  {Ye} C.~S.,   {Rasio} F.~A.,  2021, \mn@doi [\apjl] {10.3847/2041-8213/abdf5b}, \href {https://ui.adsabs.harvard.edu/abs/2021ApJ...908L..29G} {908, L29}

\bibitem[\protect\citeauthoryear{{Graham}, {Driver}, {Petrosian}, {Conselice}, {Bershady}, {Crawford}  \& {Goto}}{{Graham} et~al.}{2005}]{Graham_2005}
{Graham} A.~W.,  {Driver} S.~P.,  {Petrosian} V.,  {Conselice} C.~J.,  {Bershady} M.~A.,  {Crawford} S.~M.,   {Goto} T.,  2005, \mn@doi [\aj] {10.1086/444475}, \href {https://ui.adsabs.harvard.edu/abs/2005AJ....130.1535G} {130, 1535}

\bibitem[\protect\citeauthoryear{{Graur}, {French}, {Zahid}, {Guillochon}, {Mandel}, {Auchettl}  \& {Zabludoff}}{{Graur} et~al.}{2018}]{Graur_2018}
{Graur} O.,  {French} K.~D.,  {Zahid} H.~J.,  {Guillochon} J.,  {Mandel} K.~S.,  {Auchettl} K.,   {Zabludoff} A.~I.,  2018, \mn@doi [\apj] {10.3847/1538-4357/aaa3fd}, \href {https://ui.adsabs.harvard.edu/abs/2018ApJ...853...39G} {853, 39}

\bibitem[\protect\citeauthoryear{{Greene}, {Strader}  \& {Ho}}{{Greene} et~al.}{2020}]{Greene_2020}
{Greene} J.~E.,  {Strader} J.,   {Ho} L.~C.,  2020, \mn@doi [\araa] {10.1146/annurev-astro-032620-021835}, \href {https://ui.adsabs.harvard.edu/abs/2020ARA&A..58..257G} {58, 257}

\bibitem[\protect\citeauthoryear{{Guillochon} \& {Ramirez-Ruiz}}{{Guillochon} \& {Ramirez-Ruiz}}{2013}]{Guillochon_2013}
{Guillochon} J.,  {Ramirez-Ruiz} E.,  2013, \mn@doi [\apj] {10.1088/0004-637X/767/1/25}, \href {https://ui.adsabs.harvard.edu/abs/2013ApJ...767...25G} {767, 25}

\bibitem[\protect\citeauthoryear{{Hammerstein} et~al.,}{{Hammerstein} et~al.}{2023}]{Hammerstein_2023}
{Hammerstein} E.,  et~al., 2023, \mn@doi [\apj] {10.3847/1538-4357/aca283}, \href {https://ui.adsabs.harvard.edu/abs/2023ApJ...942....9H} {942, 9}

\bibitem[\protect\citeauthoryear{{Harris} \& {van den Bergh}}{{Harris} \& {van den Bergh}}{1981}]{Harris_1981}
{Harris} W.~E.,  {van den Bergh} S.,  1981, \mn@doi [\aj] {10.1086/113047}, \href {https://ui.adsabs.harvard.edu/abs/1981AJ.....86.1627H} {86, 1627}

\bibitem[\protect\citeauthoryear{{Hung} et~al.,}{{Hung} et~al.}{2021}]{Hung_2021}
{Hung} T.,  et~al., 2021, \mn@doi [\apj] {10.3847/1538-4357/abf4c3}, \href {https://ui.adsabs.harvard.edu/abs/2021ApJ...917....9H} {917, 9}

\bibitem[\protect\citeauthoryear{{Ivezi{\'c}} et~al.,}{{Ivezi{\'c}} et~al.}{2019}]{Ivezic_2019}
{Ivezi{\'c}} {\v{Z}}.,  et~al., 2019, \mn@doi [\apj] {10.3847/1538-4357/ab042c}, \href {https://ui.adsabs.harvard.edu/abs/2019ApJ...873..111I} {873, 111}

\bibitem[\protect\citeauthoryear{{Janz}, {Forbes}, {Norris}, {Strader}, {Penny}, {Fagioli}  \& {Romanowsky}}{{Janz} et~al.}{2015}]{Janz_2015}
{Janz} J.,  {Forbes} D.~A.,  {Norris} M.~A.,  {Strader} J.,  {Penny} S.~J.,  {Fagioli} M.,   {Romanowsky} A.~J.,  2015, \mn@doi [\mnras] {10.1093/mnras/stv389}, \href {https://ui.adsabs.harvard.edu/abs/2015MNRAS.449.1716J} {449, 1716}

\bibitem[\protect\citeauthoryear{{Janz} et~al.,}{{Janz} et~al.}{2016}]{Janz_2016}
{Janz} J.,  et~al., 2016, \mn@doi [\mnras] {10.1093/mnras/stv2636}, \href {https://ui.adsabs.harvard.edu/abs/2016MNRAS.456..617J} {456, 617}

\bibitem[\protect\citeauthoryear{{Jord{\'a}n} et~al.,}{{Jord{\'a}n} et~al.}{2007}]{Jordan_2007}
{Jord{\'a}n} A.,  et~al., 2007, \mn@doi [\apjs] {10.1086/516840}, \href {https://ui.adsabs.harvard.edu/abs/2007ApJS..171..101J} {171, 101}

\bibitem[\protect\citeauthoryear{{Kochanek} et~al.,}{{Kochanek} et~al.}{2017}]{Kochanek_2017}
{Kochanek} C.~S.,  et~al., 2017, \mn@doi [\pasp] {10.1088/1538-3873/aa80d9}, \href {https://ui.adsabs.harvard.edu/abs/2017PASP..129j4502K} {129, 104502}

\bibitem[\protect\citeauthoryear{{Law-Smith}, {Ramirez-Ruiz}, {Ellison}  \& {Foley}}{{Law-Smith} et~al.}{2017}]{Law-Smith_2017}
{Law-Smith} J.,  {Ramirez-Ruiz} E.,  {Ellison} S.~L.,   {Foley} R.~J.,  2017, \mn@doi [\apj] {10.3847/1538-4357/aa94c7}, \href {https://ui.adsabs.harvard.edu/abs/2017ApJ...850...22L} {850, 22}

\bibitem[\protect\citeauthoryear{{Law} et~al.,}{{Law} et~al.}{2009}]{Law_2009}
{Law} N.~M.,  et~al., 2009, \mn@doi [\pasp] {10.1086/648598}, \href {https://ui.adsabs.harvard.edu/abs/2009PASP..121.1395L} {121, 1395}

\bibitem[\protect\citeauthoryear{{L{\"u}tzgendorf} et~al.,}{{L{\"u}tzgendorf} et~al.}{2013}]{Lutzgendorf_2013}
{L{\"u}tzgendorf} N.,  et~al., 2013, \mn@doi [\aap] {10.1051/0004-6361/201321183}, \href {https://ui.adsabs.harvard.edu/abs/2013A&A...555A..26L} {555, A26}

\bibitem[\protect\citeauthoryear{{Makarov}, {Prugniel}, {Terekhova}, {Courtois}  \& {Vauglin}}{{Makarov} et~al.}{2014}]{Makarov_2014}
{Makarov} D.,  {Prugniel} P.,  {Terekhova} N.,  {Courtois} H.,   {Vauglin} I.,  2014, \mn@doi [\aap] {10.1051/0004-6361/201423496}, \href {https://ui.adsabs.harvard.edu/abs/2014A&A...570A..13M} {570, A13}

\bibitem[\protect\citeauthoryear{{Malyali}, {Rau}  \& {Nandra}}{{Malyali} et~al.}{2019}]{Malyali_2019}
{Malyali} A.,  {Rau} A.,   {Nandra} K.,  2019, \mn@doi [\mnras] {10.1093/mnras/stz2520}, \href {https://ui.adsabs.harvard.edu/abs/2019MNRAS.489.5413M} {489, 5413}

\bibitem[\protect\citeauthoryear{{Maraston}}{{Maraston}}{1998}]{Maraston_1998}
{Maraston} C.,  1998, \mn@doi [\mnras] {10.1046/j.1365-8711.1998.01947.x}, \href {https://ui.adsabs.harvard.edu/abs/1998MNRAS.300..872M} {300, 872}

\bibitem[\protect\citeauthoryear{{Maraston}}{{Maraston}}{2005}]{Maraston_2005}
{Maraston} C.,  2005, \mn@doi [\mnras] {10.1111/j.1365-2966.2005.09270.x}, \href {https://ui.adsabs.harvard.edu/abs/2005MNRAS.362..799M} {362, 799}

\bibitem[\protect\citeauthoryear{{Masci} et~al.,}{{Masci} et~al.}{2017}]{Masci_2017}
{Masci} F.~J.,  et~al., 2017, \mn@doi [\pasp] {10.1088/1538-3873/129/971/014002}, \href {https://ui.adsabs.harvard.edu/abs/2017PASP..129a4002M} {129, 014002}

\bibitem[\protect\citeauthoryear{{McDonald}}{{McDonald}}{2009}]{McDonald_2009}
{McDonald} I.,  2009, PhD thesis, Keele University, UK

\bibitem[\protect\citeauthoryear{{Metzger} \& {Stone}}{{Metzger} \& {Stone}}{2016}]{Metzger_2016}
{Metzger} B.~D.,  {Stone} N.~C.,  2016, \mn@doi [\mnras] {10.1093/mnras/stw1394}, \href {https://ui.adsabs.harvard.edu/abs/2016MNRAS.461..948M} {461, 948}

\bibitem[\protect\citeauthoryear{{Miller} \& {Hamilton}}{{Miller} \& {Hamilton}}{2002}]{Miller_2002}
{Miller} M.~C.,  {Hamilton} D.~P.,  2002, \mn@doi [\mnras] {10.1046/j.1365-8711.2002.05112.x}, \href {https://ui.adsabs.harvard.edu/abs/2002MNRAS.330..232C} {330, 232}

\bibitem[\protect\citeauthoryear{{Mummery}, {van Velzen}, {Nathan}, {Ingram}, {Hammerstein}, {Fraser-Taliente}  \& {Balbus}}{{Mummery} et~al.}{2024}]{Mummery_2024}
{Mummery} A.,  {van Velzen} S.,  {Nathan} E.,  {Ingram} A.,  {Hammerstein} E.,  {Fraser-Taliente} L.,   {Balbus} S.,  2024, \mn@doi [\mnras] {10.1093/mnras/stad3001}, \href {https://ui.adsabs.harvard.edu/abs/2024MNRAS.527.2452M} {527, 2452}

\bibitem[\protect\citeauthoryear{{Nicholl} et~al.,}{{Nicholl} et~al.}{2020}]{Nicholl_2020}
{Nicholl} M.,  et~al., 2020, \mn@doi [\mnras] {10.1093/mnras/staa2824}, \href {https://ui.adsabs.harvard.edu/abs/2020MNRAS.499..482N} {499, 482}

\bibitem[\protect\citeauthoryear{{Norris} \& {Kannappan}}{{Norris} \& {Kannappan}}{2011}]{Norris_2011}
{Norris} M.~A.,  {Kannappan} S.~J.,  2011, \mn@doi [\mnras] {10.1111/j.1365-2966.2011.18440.x}, \href {https://ui.adsabs.harvard.edu/abs/2011MNRAS.414..739N} {414, 739}

\bibitem[\protect\citeauthoryear{{Norris} et~al.,}{{Norris} et~al.}{2014}]{Norris_2014}
{Norris} M.~A.,  et~al., 2014, \mn@doi [\mnras] {10.1093/mnras/stu1186}, \href {https://ui.adsabs.harvard.edu/abs/2014MNRAS.443.1151N} {443, 1151}

\bibitem[\protect\citeauthoryear{{Norris}, {Escudero}, {Faifer}, {Kannappan}, {Forte}  \& {van den Bosch}}{{Norris} et~al.}{2015}]{Norris_2015}
{Norris} M.~A.,  {Escudero} C.~G.,  {Faifer} F.~R.,  {Kannappan} S.~J.,  {Forte} J.~C.,   {van den Bosch} R. C.~E.,  2015, \mn@doi [\mnras] {10.1093/mnras/stv1221}, \href {https://ui.adsabs.harvard.edu/abs/2015MNRAS.451.3615N} {451, 3615}

\bibitem[\protect\citeauthoryear{{Norris}, {van de Ven}, {Kannappan}, {Schinnerer}  \& {Leaman}}{{Norris} et~al.}{2019}]{Norris_2019}
{Norris} M.~A.,  {van de Ven} G.,  {Kannappan} S.~J.,  {Schinnerer} E.,   {Leaman} R.,  2019, \mn@doi [\mnras] {10.1093/mnras/stz2096}, \href {https://ui.adsabs.harvard.edu/abs/2019MNRAS.488.5400N} {488, 5400}

\bibitem[\protect\citeauthoryear{{Panagiotou} et~al.,}{{Panagiotou} et~al.}{2023}]{Panagiotou_2023}
{Panagiotou} C.,  et~al., 2023, \mn@doi [\apjl] {10.3847/2041-8213/acc02f}, \href {https://ui.adsabs.harvard.edu/abs/2023ApJ...948L...5P} {948, L5}

\bibitem[\protect\citeauthoryear{{Pechetti} et~al.,}{{Pechetti} et~al.}{2022}]{Pechetti_2022}
{Pechetti} R.,  et~al., 2022, \mn@doi [\apj] {10.3847/1538-4357/ac339f}, \href {https://ui.adsabs.harvard.edu/abs/2022ApJ...924...48P} {924, 48}

\bibitem[\protect\citeauthoryear{{Peng} et~al.,}{{Peng} et~al.}{2008}]{Peng_2008}
{Peng} E.~W.,  et~al., 2008, \mn@doi [\apj] {10.1086/587951}, \href {https://ui.adsabs.harvard.edu/abs/2008ApJ...681..197P} {681, 197}

\bibitem[\protect\citeauthoryear{{Pfeffer}, {Hilker}, {Baumgardt}  \& {Griffen}}{{Pfeffer} et~al.}{2016}]{Pfeffer_2016}
{Pfeffer} J.,  {Hilker} M.,  {Baumgardt} H.,   {Griffen} B.~F.,  2016, \mn@doi [\mnras] {10.1093/mnras/stw498}, \href {https://ui.adsabs.harvard.edu/abs/2016MNRAS.458.2492P} {458, 2492}

\bibitem[\protect\citeauthoryear{{Portegies Zwart} \& {McMillan}}{{Portegies Zwart} \& {McMillan}}{2002}]{PortegiesZwart_2002}
{Portegies Zwart} S.~F.,  {McMillan} S. L.~W.,  2002, \mn@doi [\apj] {10.1086/341798}, \href {https://ui.adsabs.harvard.edu/abs/2002ApJ...576..899P} {576, 899}

\bibitem[\protect\citeauthoryear{{Rodriguez}, {Amaro-Seoane}, {Chatterjee}  \& {Rasio}}{{Rodriguez} et~al.}{2018}]{Rodriguez_2018}
{Rodriguez} C.~L.,  {Amaro-Seoane} P.,  {Chatterjee} S.,   {Rasio} F.~A.,  2018, \mn@doi [\prl] {10.1103/PhysRevLett.120.151101}, \href {https://ui.adsabs.harvard.edu/abs/2018PhRvL.120o1101R} {120, 151101}

\bibitem[\protect\citeauthoryear{{Rodriguez}, {Zevin}, {Amaro-Seoane}, {Chatterjee}, {Kremer}, {Rasio}  \& {Ye}}{{Rodriguez} et~al.}{2019}]{Rodriguez_2019}
{Rodriguez} C.~L.,  {Zevin} M.,  {Amaro-Seoane} P.,  {Chatterjee} S.,  {Kremer} K.,  {Rasio} F.~A.,   {Ye} C.~S.,  2019, \mn@doi [\prd] {10.1103/PhysRevD.100.043027}, \href {https://ui.adsabs.harvard.edu/abs/2019PhRvD.100d3027R} {100, 043027}

\bibitem[\protect\citeauthoryear{{Seth} et~al.,}{{Seth} et~al.}{2014}]{Seth_2014}
{Seth} A.~C.,  et~al., 2014, \mn@doi [\nat] {10.1038/nature13762}, \href {https://ui.adsabs.harvard.edu/abs/2014Natur.513..398S} {513, 398}

\bibitem[\protect\citeauthoryear{{Shen}, {Mo}, {White}, {Blanton}, {Kauffmann}, {Voges}, {Brinkmann}  \& {Csabai}}{{Shen} et~al.}{2003}]{Shen_2003}
{Shen} S.,  {Mo} H.~J.,  {White} S. D.~M.,  {Blanton} M.~R.,  {Kauffmann} G.,  {Voges} W.,  {Brinkmann} J.,   {Csabai} I.,  2003, \mn@doi [\mnras] {10.1046/j.1365-8711.2003.06740.x}, \href {https://ui.adsabs.harvard.edu/abs/2003MNRAS.343..978S} {343, 978}

\bibitem[\protect\citeauthoryear{{Stone} \& {Metzger}}{{Stone} \& {Metzger}}{2016}]{Stone_2016}
{Stone} N.~C.,  {Metzger} B.~D.,  2016, \mn@doi [\mnras] {10.1093/mnras/stv2281}, \href {https://ui.adsabs.harvard.edu/abs/2016MNRAS.455..859S} {455, 859}

\bibitem[\protect\citeauthoryear{{Strader}, {Chomiuk}, {Maccarone}, {Miller-Jones}  \& {Seth}}{{Strader} et~al.}{2012}]{Strader_2012a}
{Strader} J.,  {Chomiuk} L.,  {Maccarone} T.~J.,  {Miller-Jones} J. C.~A.,   {Seth} A.~C.,  2012, \mn@doi [\nat] {10.1038/nature11490}, \href {https://ui.adsabs.harvard.edu/abs/2012Natur.490...71S} {490, 71}

\bibitem[\protect\citeauthoryear{{Tang}, {Madau}, {Bortolas}, {Peng}, {Feng}  \& {Guhathakurta}}{{Tang} et~al.}{2024}]{Tang_2024}
{Tang} V.~L.,  {Madau} P.,  {Bortolas} E.,  {Peng} E.~W.,  {Feng} Y.,   {Guhathakurta} P.,  2024, \mn@doi [\apj] {10.3847/1538-4357/ad1dd9}, \href {https://ui.adsabs.harvard.edu/abs/2024ApJ...963..146T} {963, 146}

\bibitem[\protect\citeauthoryear{{Tonry} et~al.,}{{Tonry} et~al.}{2018}]{Tonry_2018}
{Tonry} J.~L.,  et~al., 2018, \mn@doi [\pasp] {10.1088/1538-3873/aabadf}, \href {https://ui.adsabs.harvard.edu/abs/2018PASP..130f4505T} {130, 064505}

\bibitem[\protect\citeauthoryear{{Udalski}, {Szyma{\'n}ski}  \& {Szyma{\'n}ski}}{{Udalski} et~al.}{2015}]{Udalski_2015}
{Udalski} A.,  {Szyma{\'n}ski} M.~K.,   {Szyma{\'n}ski} G.,  2015, \actaa, \href {https://ui.adsabs.harvard.edu/abs/2015AcA....65....1U} {65, 1}

\bibitem[\protect\citeauthoryear{{Ulmer}}{{Ulmer}}{1999}]{Ulmer_1999}
{Ulmer} A.,  1999, \mn@doi [\apj] {10.1086/306909}, \href {https://ui.adsabs.harvard.edu/abs/1999ApJ...514..180U} {514, 180}

\bibitem[\protect\citeauthoryear{{Vitral} \& {Mamon}}{{Vitral} \& {Mamon}}{2021}]{Vitral_2021}
{Vitral} E.,  {Mamon} G.~A.,  2021, \mn@doi [\aap] {10.1051/0004-6361/202039650}, \href {https://ui.adsabs.harvard.edu/abs/2021A&A...646A..63V} {646, A63}

\bibitem[\protect\citeauthoryear{{Voggel} et~al.,}{{Voggel} et~al.}{2018}]{Voggel_2018}
{Voggel} K.~T.,  et~al., 2018, \mn@doi [\apj] {10.3847/1538-4357/aabae5}, \href {https://ui.adsabs.harvard.edu/abs/2018ApJ...858...20V} {858, 20}

\bibitem[\protect\citeauthoryear{{Wang} \& {Merritt}}{{Wang} \& {Merritt}}{2004}]{Wang_2004}
{Wang} J.,  {Merritt} D.,  2004, \mn@doi [\apj] {10.1086/379767}, \href {https://ui.adsabs.harvard.edu/abs/2004ApJ...600..149W} {600, 149}

\bibitem[\protect\citeauthoryear{{Wenger} et~al.,}{{Wenger} et~al.}{2000}]{Wenger_2000}
{Wenger} M.,  et~al., 2000, \mn@doi [\aaps] {10.1051/aas:2000332}, \href {https://ui.adsabs.harvard.edu/abs/2000A&AS..143....9W} {143, 9}

\bibitem[\protect\citeauthoryear{{Yao} et~al.,}{{Yao} et~al.}{2023}]{Yao_2023}
{Yao} Y.,  et~al., 2023, \mn@doi [arXiv e-prints] {10.48550/arXiv.2303.06523}, \href {https://ui.adsabs.harvard.edu/abs/2023arXiv230306523Y} {p. arXiv:2303.06523}

\bibitem[\protect\citeauthoryear{{de Vaucouleurs}, {de Vaucouleurs}, {Corwin}, {Buta}, {Paturel}  \& {Fouque}}{{de Vaucouleurs} et~al.}{1991}]{deVaucouleurs_1991}
{de Vaucouleurs} G.,  {de Vaucouleurs} A.,  {Corwin} Herold~G. J.,  {Buta} R.~J.,  {Paturel} G.,   {Fouque} P.,  1991, {Third Reference Catalogue of Bright Galaxies}.
Springer, New York, NY (USA)

\bibitem[\protect\citeauthoryear{{van Velzen} et~al.,}{{van Velzen} et~al.}{2011}]{vanVelzen_2011}
{van Velzen} S.,  et~al., 2011, \mn@doi [\apj] {10.1088/0004-637X/741/2/73}, \href {https://ui.adsabs.harvard.edu/abs/2011ApJ...741...73V} {741, 73}

\bibitem[\protect\citeauthoryear{{van Velzen}, {Holoien}, {Onori}, {Hung}  \& {Arcavi}}{{van Velzen} et~al.}{2020}]{vanVelzen_2020}
{van Velzen} S.,  {Holoien} T. W.~S.,  {Onori} F.,  {Hung} T.,   {Arcavi} I.,  2020, \mn@doi [\ssr] {10.1007/s11214-020-00753-z}, \href {https://ui.adsabs.harvard.edu/abs/2020SSRv..216..124V} {216, 124}

\bibitem[\protect\citeauthoryear{{van Velzen} et~al.,}{{van Velzen} et~al.}{2021}]{vanVelzen_2021}
{van Velzen} S.,  et~al., 2021, \mn@doi [\apj] {10.3847/1538-4357/abc258}, \href {https://ui.adsabs.harvard.edu/abs/2021ApJ...908....4V} {908, 4}

\bibitem[\protect\citeauthoryear{{van den Bergh}}{{van den Bergh}}{2006}]{vandenBergh_2006}
{van den Bergh} S.,  2006, \mn@doi [\aj] {10.1086/498688}, \href {https://ui.adsabs.harvard.edu/abs/2006AJ....131..304V} {131, 304}

\makeatother
\end{thebibliography}




{\appendix

\section{Light curve categorisation}\label{Light curve categorisation}
As noted in the main text, visual classification of 1900 ZTF alert object light curves was carried out after initial filtering of the alerts. The categories were defined to facilitate an iterative process of object reduction, so that further scrutiny could be prioritised against objects of interest. As shown in the unclassified transients section of \cref{tab:tns_summary} these categories were defined as spurious, nuclear, spiral arm / halo located, background galaxy, imaging artefacts, stellar object, positive template. The final category was indeterminate, for objects which defied classification in any of the previous bins. Examples of each of these categories are shown in \cref{fig:LC_medley}.

\begin{figure*}
    \centering
    \begin{minipage}{.48\linewidth}
        \centering
        \subcaptionbox{One of the 2 TDEs observed in the sample. Both were nuclear TDEs (non-halo) - included for reference purposes.}
            {\includegraphics[width=0.78\linewidth]{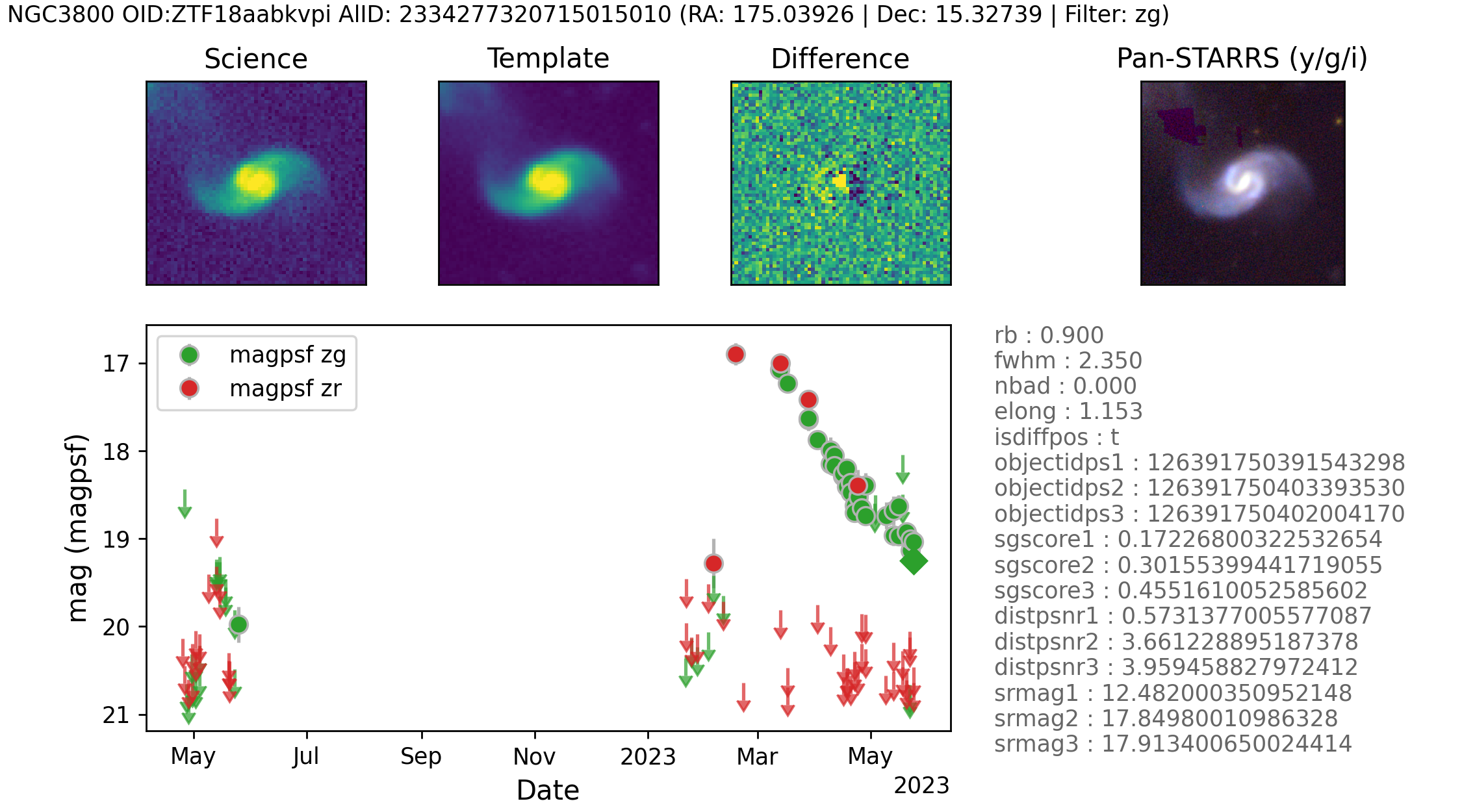}}
        \vspace{5pt}
        \subcaptionbox{Typical spurious alert. Minimal uncorrelated detections with multiple non-detections.}
            {\includegraphics[width=0.78\linewidth]{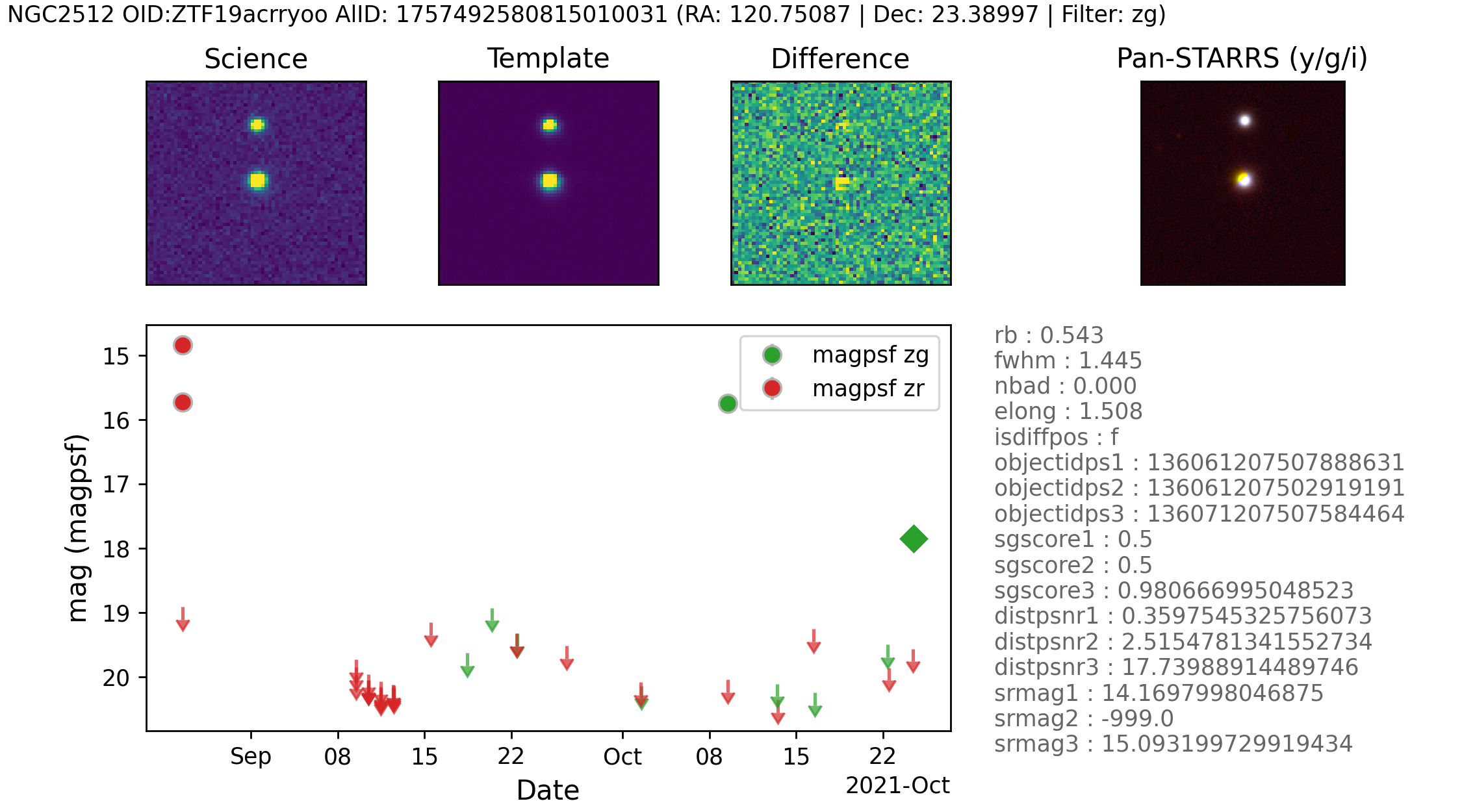}}
        \vspace{5pt}
        \subcaptionbox{Alerts in proximity to host galaxy nucleus.}
            {\includegraphics[width=0.78\linewidth]{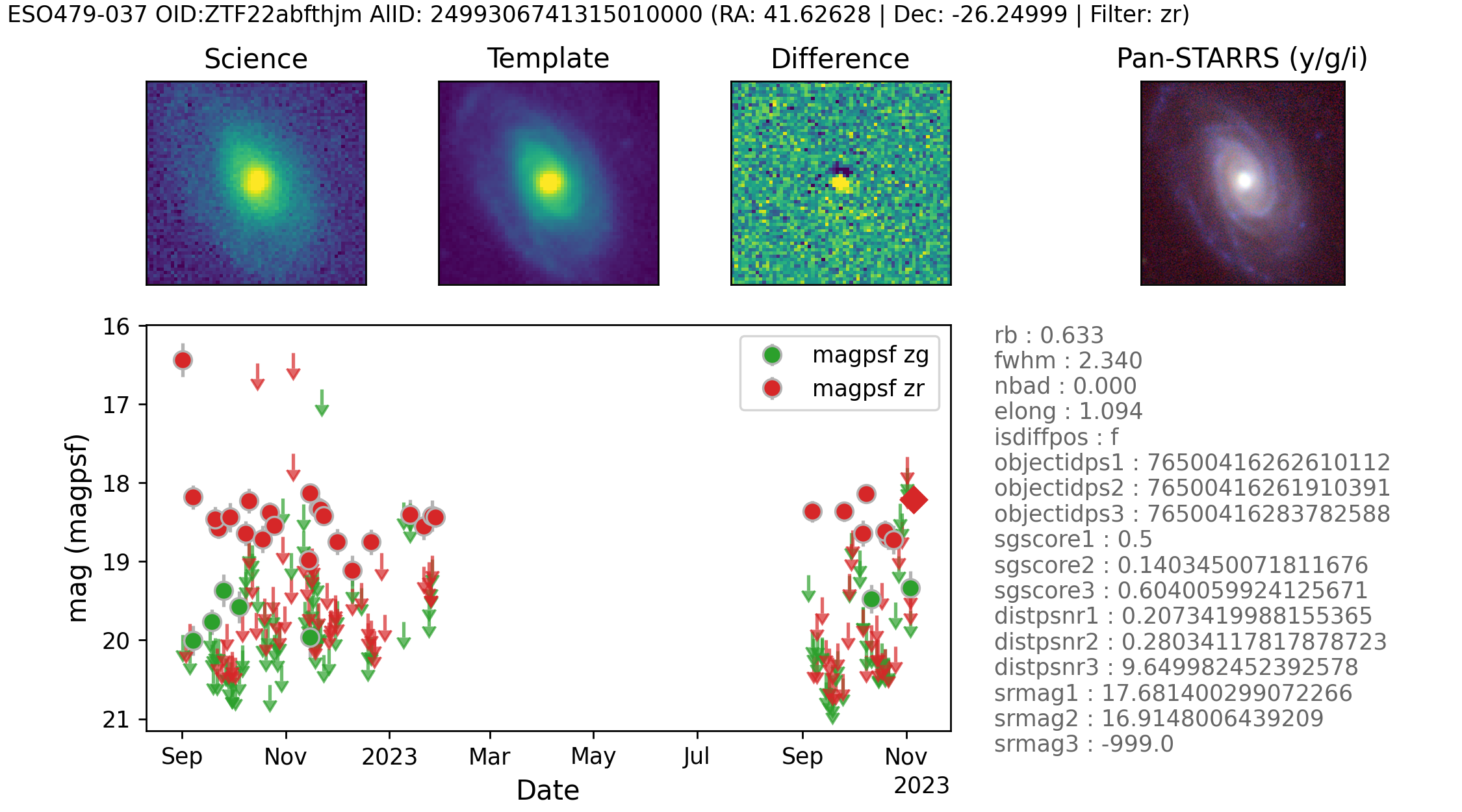}}
        \vspace{5pt}
        \subcaptionbox{Typical spiral arm `clutter' causing spurious alerts.}
            {\includegraphics[width=0.78\linewidth]{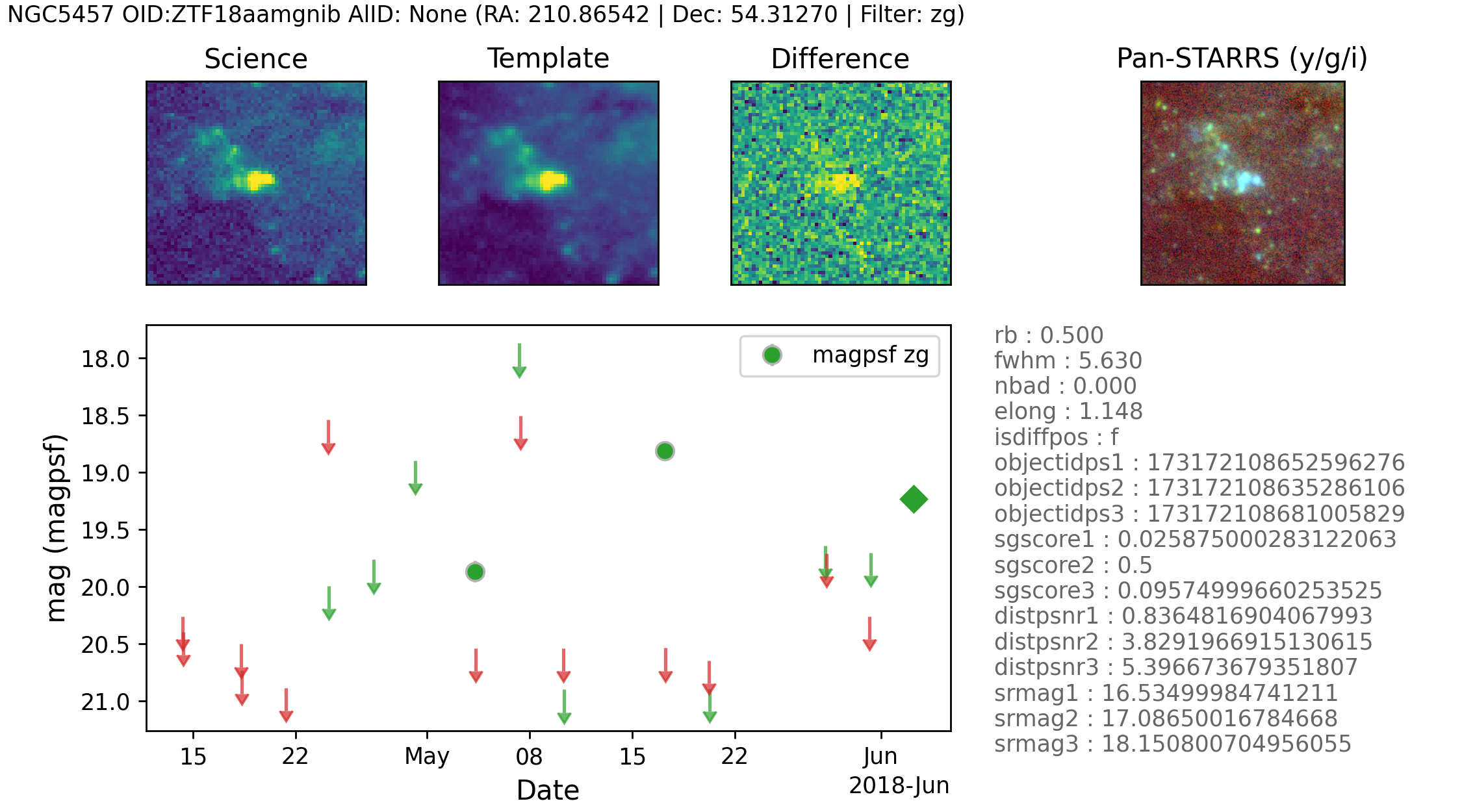}}
        \vspace{5pt}
        \subcaptionbox{Halo detection. As noted in the text, similar objects were given more scrutiny. This object rejected as unclassified SNe-type II, based on rise time and colour evolution.}
            {\includegraphics[width=0.78\linewidth]{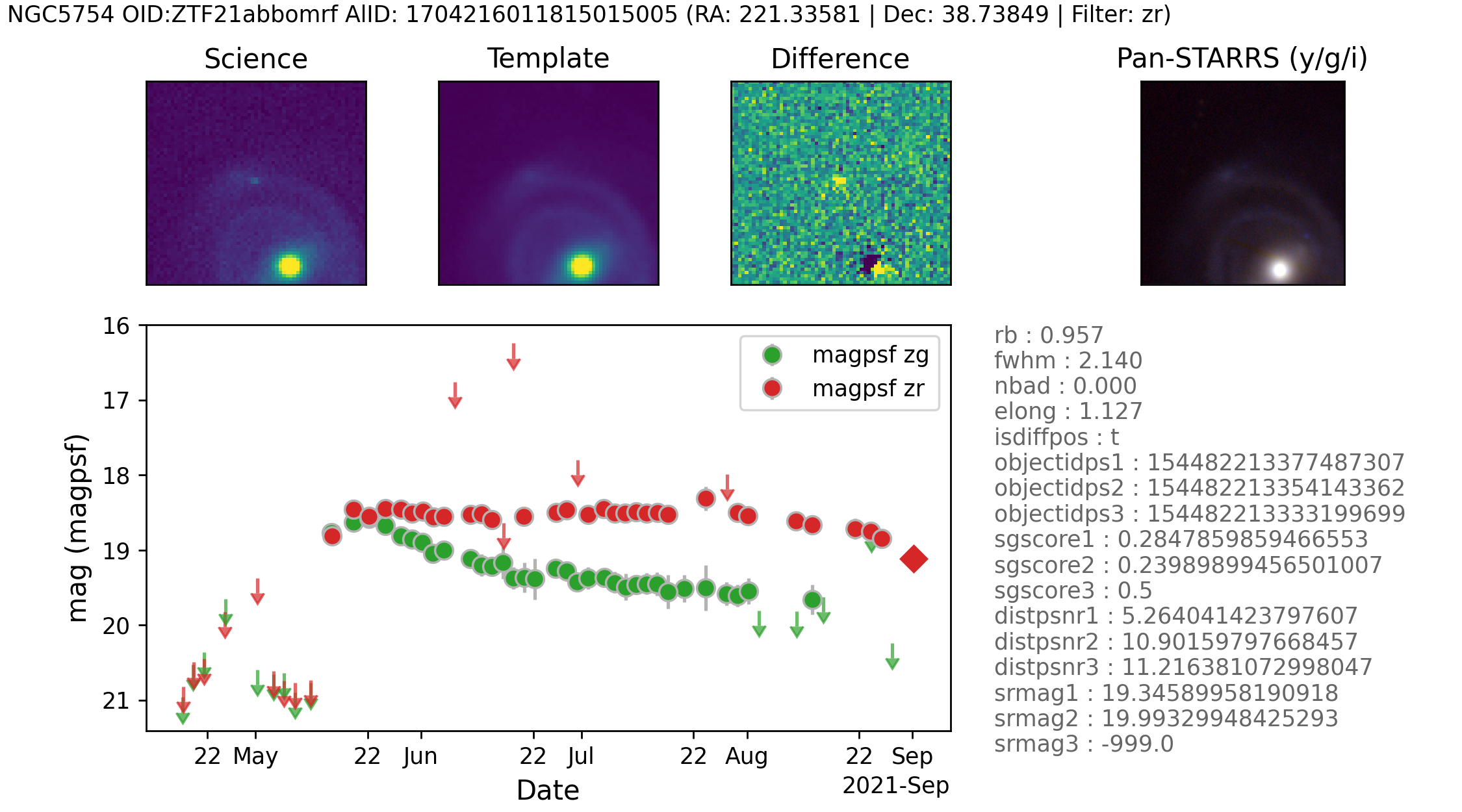}}
    \end{minipage} 
    \begin{minipage}{.45\linewidth}
        \centering
        \subcaptionbox{Imaging artefact associated with stellar object.}
            {\includegraphics[width=0.78\linewidth]{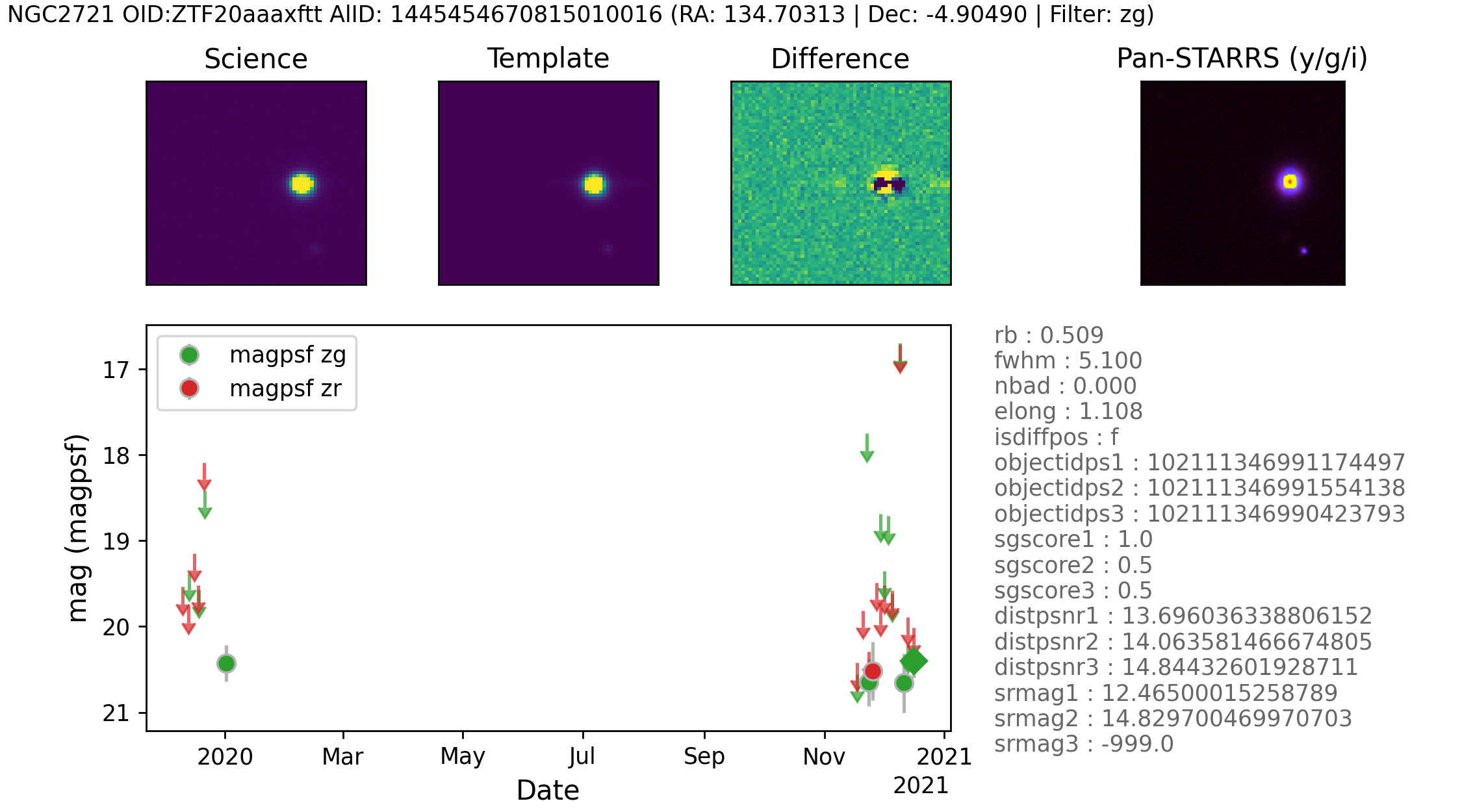}}
        \vspace{5pt}
        \subcaptionbox{Indeterminate alerting detections. No distinct curve, multiple intermediate non-detections.}
            {\includegraphics[width=0.78\linewidth]{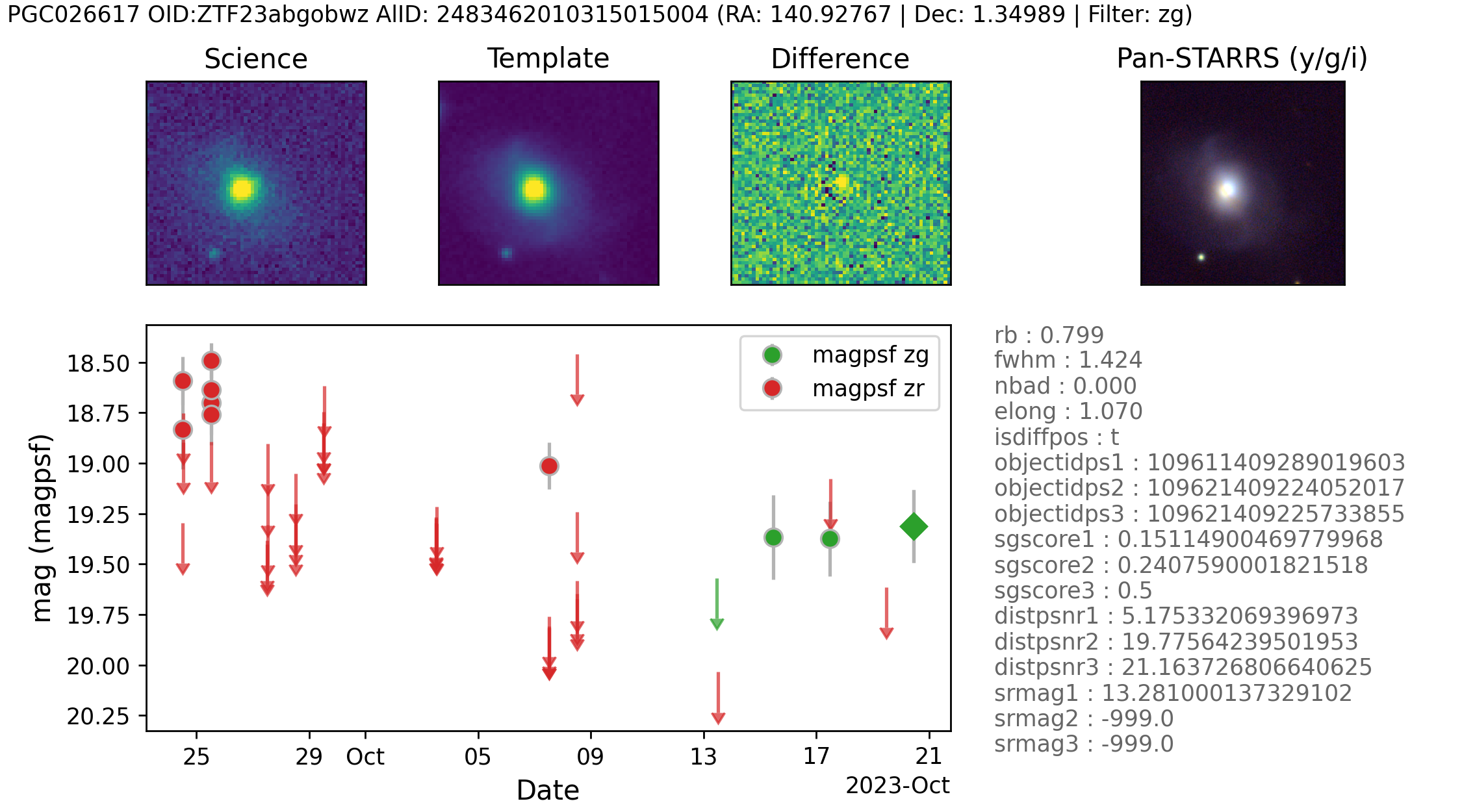}}
        \vspace{5pt}
        \subcaptionbox{Objects associated with `local' variable star or stellar variation.}
            {\includegraphics[width=0.78\linewidth]{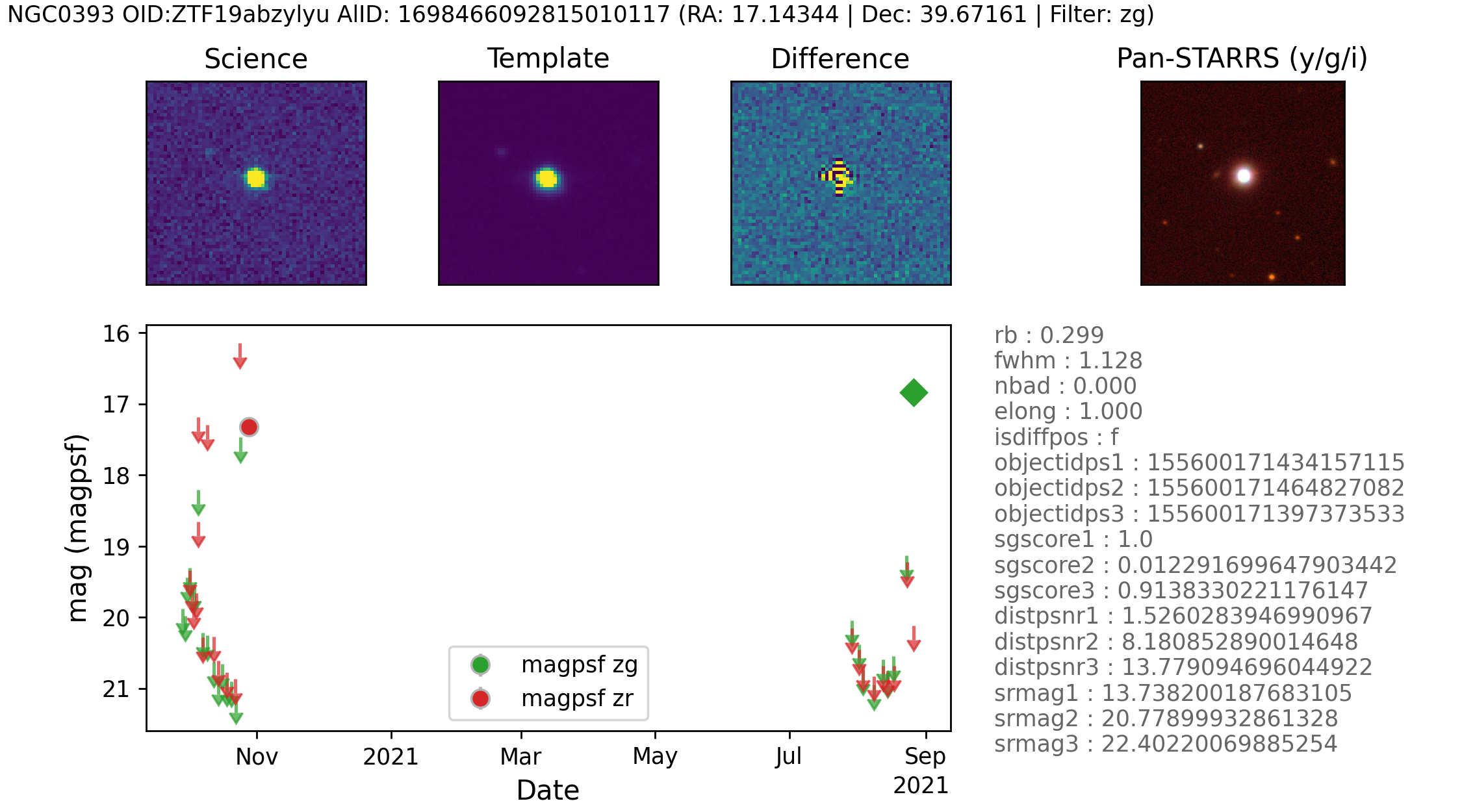}}
        \vspace{5pt}
        \subcaptionbox{Multiple alerts showing negative difference based on legacy positive template (i.e. transient during reference image generation).}
            {\includegraphics[width=0.78\linewidth]{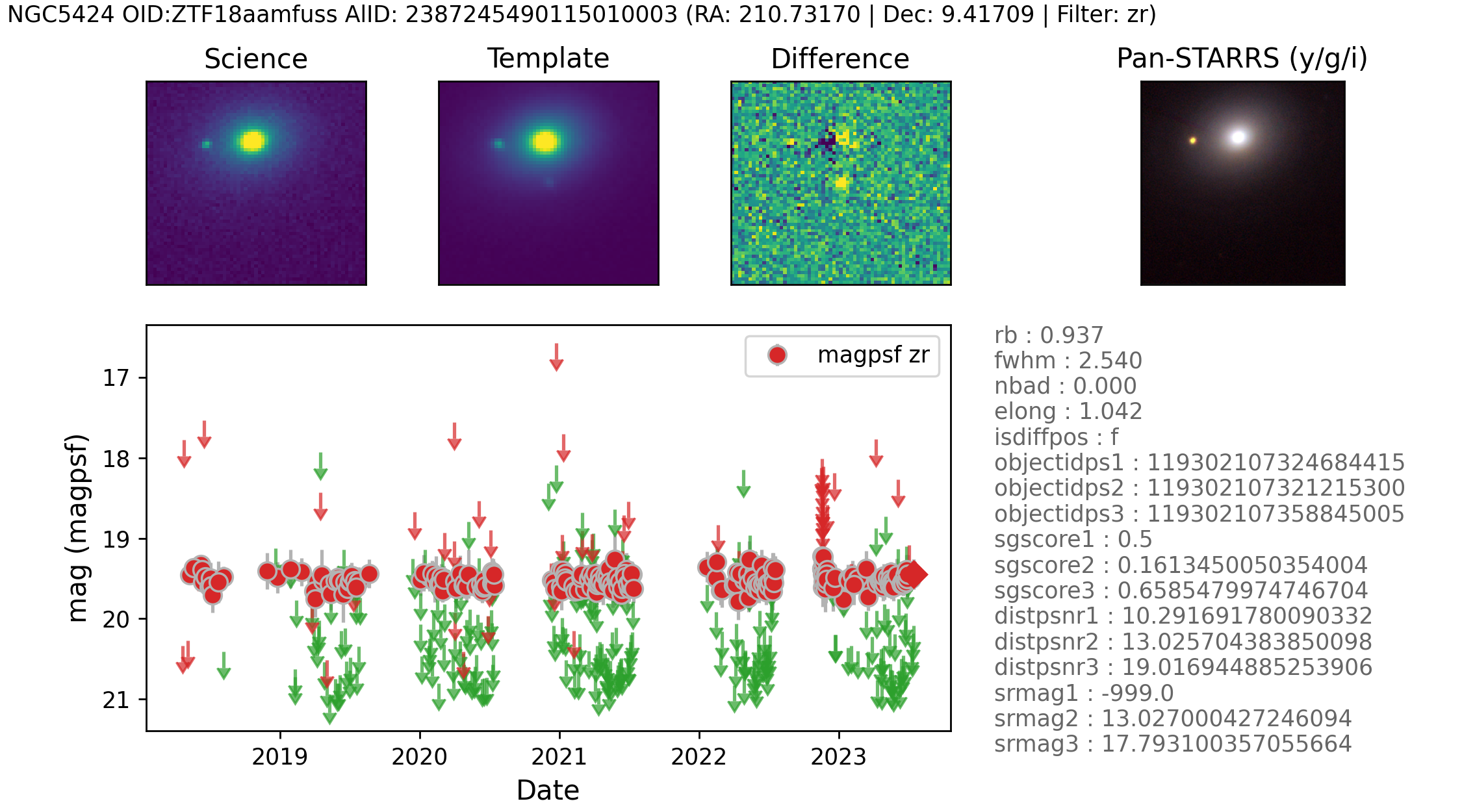}}
        \vspace{5pt}
        \subcaptionbox{Background galaxy or AGN associated alerts.}
            {\includegraphics[width=0.78\linewidth]{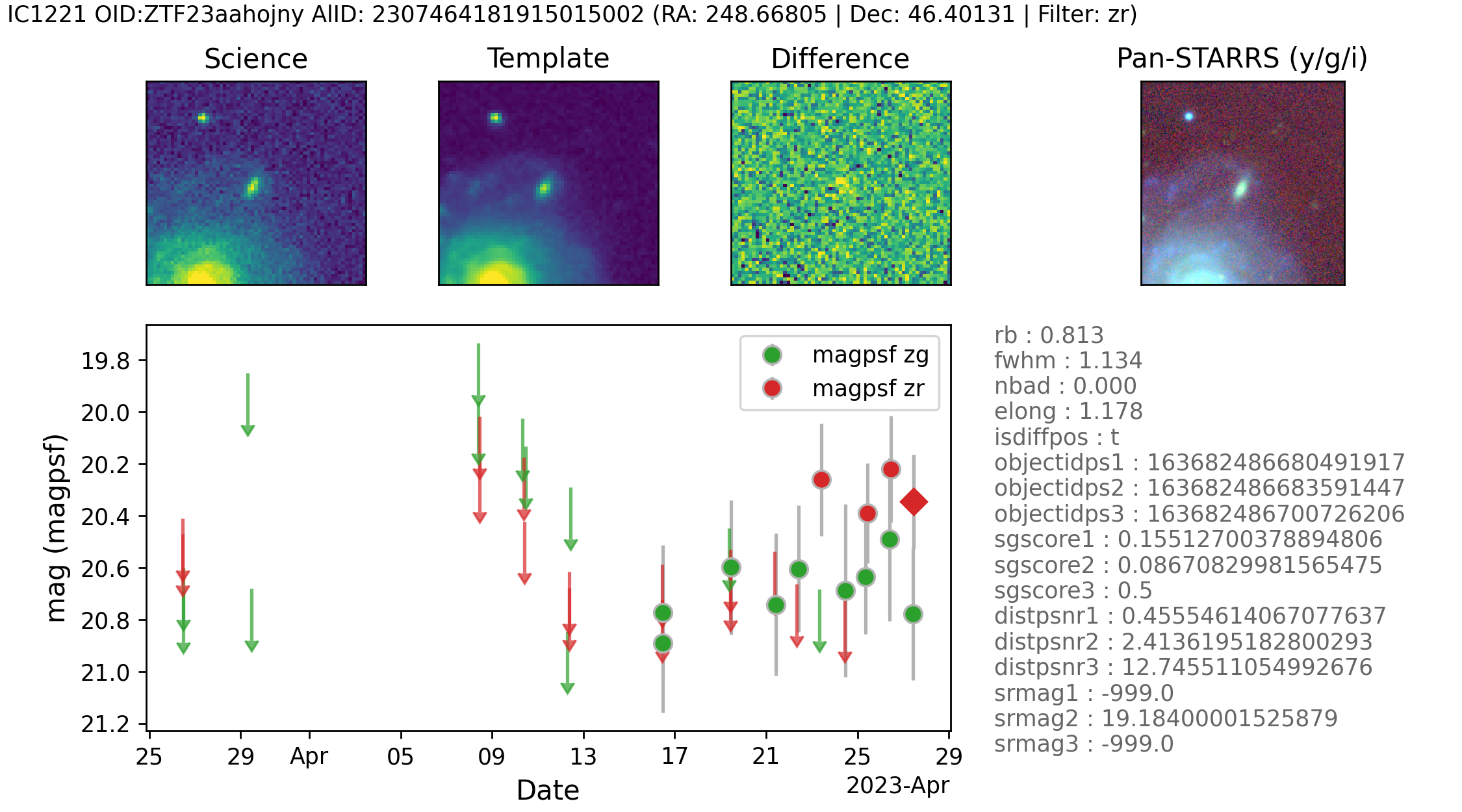}}
    
    \end{minipage}
    \caption{Example lightcurves, illustrating different categories (as detailed in main text) used during the ZTF alert object classification process.}
    \label{fig:LC_medley}

\end{figure*}


\bsp	
\label{lastpage}
\end{document}